\documentclass[%
reprint,
superscriptaddress,
preprintnumbers,
nofootinbib,
nobibnotes,
amsmath,amssymb,
aps,
prc,
]{revtex4-2}
\usepackage[pdftex]{graphicx}
\usepackage{dcolumn}
\usepackage{bm}
\usepackage[pdftex]{color}
\usepackage{CJK}
\usepackage[T1]{fontenc}
\usepackage{mathrsfs}
\def\ve#1{{\bm{#1}}}
\def\nuc#1#2#3{{}^{#2}_{#3}\mathrm{#1}}
\def\urm#1{\scriptstyle{\text{\textrm{\textmd{\textup{#1}}}}}}
\def\uurm#1{\scriptscriptstyle{\text{\textrm{\textmd{\textup{#1}}}}}}
\def\avr#1{\left\langle{#1}\right\rangle}

\def\ca#1{{\mathcal{#1}}}

\def\FaNDF{$ \text{FaNDF}^{\text{0}} $}
\let\temp\epsilon
\let\epsilon\varepsilon
\let\varepsilon\temp
\let\temp\relax
\let\temp\phi
\let\phi\varphi
\let\varphi\temp
\let\temp\relax
\DeclareMathOperator{\laplace}{\Delta}
\begin{document}
%
\begin{CJK*}{UTF8}{}
  \preprint{RIKEN-iTHEMS-Report-26}
  \title{Microscopic mechanism underlying charge-radius trends in Fayans pairing}
  \author{Tomoya Naito (\CJKfamily{min}{内藤智也})}
  \affiliation{
    Department of Nuclear Engineering and Management, Graduate School of Engineering, The University of Tokyo,
    Tokyo 113-8656, Japan}
  \affiliation{
    RIKEN Center for Interdisciplinary Theoretical and Mathematical Sciences (iTHEMS),
    Wako 351-0198, Japan}
  \affiliation{
    RIKEN Nishina Center, Wako 351-0198, Japan}
  \author{Gianluca Col\`{o}}
  \affiliation{
    Dipartimento di Fisica, Universit\`{a} degli Studi di Milano,
    Via Celoria 16, 20133 Milano, Italy}
  \affiliation{
    INFN, Sezione di Milano,
    Via Celoria 16, 20133 Milano, Italy}
  \author{Xavier Roca-Maza}
  \affiliation{
    Departament de F\'{\i}sica Qu\`{a}ntica i Astrof\'{\i}sica,
    Universitat de Barcelona, 
    Mart\'{\i} i Franqu\'{e}s 1,
    08028 Barcelona, Spain}
  \affiliation{
    Institut de Ci\`{e}ncies del Cosmos,
    Universitat de Barcelona, 
    Mart\'{\i} i Franqu\'{e}s 1,
    08028 Barcelona, Spain}
  \affiliation{
    Dipartimento di Fisica, Universit\`{a} degli Studi di Milano,
    Via Celoria 16, 20133 Milano, Italy}
  \affiliation{
    INFN, Sezione di Milano,
    Via Celoria 16, 20133 Milano, Italy}
  \author{Hiroyuki Sagawa (\CJKfamily{min}{佐川弘幸})}
  \affiliation{
    Center for Mathematics and Physics, University of Aizu,
    Aizu-Wakamatsu 965-8560, Japan}
  \affiliation{
    RIKEN Nishina Center, Wako 351-0198, Japan}
  \author{Enrico Vigezzi}
  \affiliation{
    INFN, Sezione di Milano,
    Via Celoria 16, 20133 Milano, Italy}
  \date{\today}
  \begin{abstract}
    The Fayans energy density functional (EDF), and in particular its pairing sector,
    has been claimed to be able to reproduce the experimental data of charge radii in many instances.
    A particularly intriguing case is that of the $ \mathrm{Ca} $ isotopes between $ A = 40 $ and $ 48 $,
    where charge radii exhibit a ``bell shape.'' 
    In our work, we examine the microscopic origin of this behavior.
    The pairing part of the Fayans EDF contains three essential parameters:
    the overall strength $ f $
    and the density-dependent and density-gradient terms, $ \tilde{h}_0 $ and $ \tilde{h}_{\urm{D}} $.
    We select a number of parametrizations,
    fixing the overall strength $ f $ to reproduce the pairing gap of $ {}^{44} \mathrm{Ca} $.
    We find that some parameter sets enhance the charge radii in open-shell nuclei
    and reproduce the bell shape of the charge radii of $ \mathrm{Ca} $ isotopes.
    It is found that this effect originates from the repulsive nature of the rearrangement potential,
    and cannot be mocked up by a refit of the pairing strength.
    On the other hand, none of the parametrizations that we have considered can reproduce satisfactorily the experimental charge radii of
    $ \mathrm{Ca} $, $ \mathrm{Sn} $, and $ \mathrm{Pb} $ isotopes at the same time. 
  \end{abstract}
  \maketitle
\end{CJK*}
%
%
\section{Introduction}
\label{sec:intro}
\par
The charge radius of an atomic nucleus is one of the physical observables which have been measured systematically and accurately via, for instance, electron scattering~\cite{
  DeVries1987At.DataNucl.DataTables36_495,
  Tsukada2017Phys.Rev.Lett.118_262501,
  Tsukada2023Phys.Rev.Lett.131_092502},
laser spectroscopy~\cite{
  Angeli2013At.DataNucl.DataTables99_69,
  GarciaRuiz2016Nat.Phys.12_594,
  Raeder2018Phys.Rev.Lett.120_232503,
  Miller2019Nat.Phys.15_432,
  Gorges2019Phys.Rev.Lett.122_192502,
  Koszorus2021Nat.Phys.17_439,
  DayGoodacre2021Phys.Rev.Lett.126_032502,
  Yang2023Prog.Part.Nucl.Phys.129_104005,
  Warbinek2024Nature634_1075},
and muonic x-rays~\cite{
  Xie2023Phys.Lett.B846_138232,
  Xie2024Phys.Rev.C109_034309,
  Saito2025Phys.Rev.C111_034313}.
Recently, the charge radii of proton-rich $ \mathrm{Ca} $ isotopes~\cite{
  Miller2019Nat.Phys.15_432},
neutron-rich
$ \mathrm{K} $~\cite{
  Koszorus2021Nat.Phys.17_439},
$ \mathrm{Ca} $~\cite{
  GarciaRuiz2016Nat.Phys.12_594},
$ \mathrm{Sn} $~\cite{
  Gorges2019Phys.Rev.Lett.122_192502},
and 
$ \mathrm{Hg} $~\cite{
  DayGoodacre2021Phys.Rev.Lett.126_032502}
isotopes,
and superheavy nuclei such as $ \mathrm{Fm} $~\cite{
  Warbinek2024Nature634_1075}
and $ \mathrm{No} $~\cite{
  Raeder2018Phys.Rev.Lett.120_232503}
have been measured. 
Together with the mass, or the binding energy, of a nucleus, the charge radius is
a fundamental observable that represents a primary benchmark for any nuclear structure model,
before addressing more subtle properties. 
Radii are sensitive to shell structure, weak nuclear binding, and pairing effects.
\par
In this paper, 
we apply nuclear density functional theory (DFT)~\cite{
  Schunck2019EnergyDensityFunctionalMethodsforAtomicNuclei_IOPPublishing,
  Colo2020Adv.Phys.X5_1740061}
to the study of charge radii. 
Nuclear DFT can be applied throughout the whole isotope chart,
and bulk properties, such as nuclear binding energies, radii, and deformations,
are among the most basic properties on which one can judge the reliability of a specific energy density functional (EDF).
Charge radii obtained by DFT are, basically, in good agreement with the experimental data~\cite{
  Roca-Maza2018Prog.Part.Nucl.Phys.101_96,
  Perera2021Phys.Rev.C104_064313,
  Naito2023Phys.Rev.C107_054307},
yet with exceptions.
In the following, we concentrate on the isotopic behavior of charge radii in spherical, open-shell nuclei.
First of all, the EDFs should reproduce the global trends of the odd-even mass differences over the mass table.
There are open problems concerning different observables:
\par
(1) The ``kink'' behavior of the isotope shift~\cite{
  Reinhard1995Nucl.Phys.A584_467,
  Nakada2019Phys.Rev.C100_044310,
  Perera2021Phys.Rev.C104_064313,
  Naito2023Phys.Rev.C107_054307,
  Marcos2024Eur.Phys.J.A60_78,
  Marcos2024Nucl.Phys.A1047_122883},
which is the change in the slope of the charge radii as a function of the mass number $ A $ below and above the magic numbers;
this is related to the symmetry energy as one goes toward neutron-rich or proton-rich nuclei,
as well as to the spin-orbit and pairing interactions that govern the detailed position of the levels and their occupation probabilities.
\par
(2) The isotopic dependence of the charge radii of $ \mathrm{Ca} $ isotopes, even the stable ones~\cite{
  Miyagi2020Phys.Rev.C102_034320,
  Perera2021Phys.Rev.C104_064313,
  Naito2023Phys.Rev.C107_054307,
  Heinz2025Phys.Rev.C111_034311},
represents a specific and difficult open question. 
The charge radii of most proton-magic nuclei display a constant increase with respect to the mass number $ A $.
In contrast, the charge radii of the two closed-shell nuclei $ \nuc{Ca}{40}{} $ and $ \nuc{Ca}{48}{} $ are almost the same, 
and that of $ \nuc{Ca}{44}{} $ is larger.
Conventional theoretical approaches for nuclear structure, including most EDFs, cannot reproduce 
this ``bell shape'' behavior of the charge radii of $ \mathrm{Ca} $ isotopes~\cite{
  Bender2003Rev.Mod.Phys.75_121}.
\par
(3) The odd-even staggering of radii, observed in various isotopic chains.
\par
The family of EDFs developed by Fayans and several collaborators over many years stands out for the good reproduction of these observables.
Moreover, it has attracted much attention because of its original form,
which includes a more complex density dependence, as compared to the most widely used Skyrme and Gogny EDFs.
This dependence produces an important rearrangement term and a stronger dependence of the radii on the particle-particle ($ p $-$ p $) channel, as compared to most EDFs.
\par
The Fayans EDF originates from Migdal's theory of finite Fermi systems~\cite{
  Khodel1982J.Phys.G8_967,
  Khodel1987Nucl.Phys.A465_397,
  Fayans1994Nucl.Phys.A568_523}.
It has a different density dependence, as compared to Skyrme or Gogny.
The form of the Fayans EDF was originally proposed in Ref.~\cite{
  Kim1992Phys.Rev.C46_1656}.
In an early phase, the work was limited to nonsuperfluid nuclei.
A set of EDFs named DF1, DF2, and DF3 was later obtained by fitting the binding energies,
the radii, and the single-particle spectra of the closed-shell nuclei $ \nuc{Ca}{40}{} $, $ \nuc{Ca}{48}{} $, and $ \nuc{Pb}{208}{} $ as well as some open-shell nuclei using a simple,
zero-range pairing interaction.
The DF3 EDF introduced in Ref.~\cite{
  Borzov1996Z.Phys.A355_117}
was particularly successful
and also used to calculate $ \beta $-decay lifetimes~\cite{
  Borzov1996Z.Phys.A355_117,
  Borzov2011Phys.At.Nucl.74_1435}.
\par
Fitting the binding energies and radii throughout the mass table calls for a density dependence of the normal part of the effective Hamiltonian;
one could expect that odd-even effects in masses and radii could provide, in turn, information on the density dependence of the pairing interaction.
It was found that the introduction of a pairing interaction which changes suddenly as a function of density,
producing a strong attraction outside of the nucleus and a nearly vanishing strength inside the volume~\cite{
  Bertsch1991Ann.Phys.209_327},
could substantially improve the reproduction of the experimental odd-even staggering and kinks observed in charge radii~\cite{
  Fayans1994Phys.Lett.B338_1}.
The special feature of the pairing interaction proposed in Ref.~\cite{
  Fayans1996Phys.Lett.B383_19}
was the introduction of a repulsive term proportional to the square of the gradient of the total nuclear density.
This interaction was employed in Ref.~\cite{
  Fayans2000Nucl.Phys.A676_49}
to calculate the odd-even staggering in $ \mathrm{Ca} $, $ \mathrm{Sn} $, and $ \mathrm{Pb} $ chains.
Satisfactory results were found, although it was necessary to increase the strength of the pairing interaction by $ 35 \, \% $ in order to reproduce the behavior in $ \mathrm{Ca} $ isotopes.
The finite-range DF3 and DF3a EDFs (the latter contains small modifications in the spin-orbit part) and some more recent versions  were then employed in a series of calculations of charge radii in different isotopic chains~\cite{
  Saperstein2016Phys.At.Nucl.79_1030,
  Borzov2022Phys.At.Nucl.85_222,
  Borzov2023Phys.Part.Nucl.54_586},
of the properties of low-lying quadrupole states~\cite{
  Tolokonnikov2011Phys.Rev.C84_064324,
  Saperstein2016Phys.At.Nucl.79_1030},
of the excitation of isobaric analog states~\cite{
  Borzov2020Phys.At.Nucl.83_24},
of beta-decay lifetimes~\cite{
  Borzov2020Phys.At.Nucl.83_700},
and 
of spin-dipole excitations~\cite{
  Borzov2024Phys.At.Nucl.87_541}.
\par
A zero-range version of the Fayans EDF, named {\FaNDF}, was proposed in Ref.~\cite{
  Fayans1998JETPLett.68_169}.
The EDF form is assumed based on the Pad\'{e} approximation,
and 
the parameters of the particle-hole ($ p $-$ h $) part
of the EDF were determined by reproducing the equation of state of neutron and nuclear matter calculated with the $ \text{UV14} + \text{TNI} $ interaction,
as well as the masses of about 100 spherical nuclei.
The pairing part of the EDF was taken from Ref.~\cite{
  Fayans1996Phys.Lett.B383_19},
as in the case of the DF3 EDF.
The zero-range form of the EDF is obviously very convenient for numerical calculations;
however, it was only used in a few cases, mostly for axially deformed calculations~\cite{
  Tolokonnikov2015J.Phys.G42_075102,
  Saperstein2015JETPLett.102_421,
  Tolokonnikov2016Phys.At.Nucl.79_21}.
The parameters of the pairing interactions were never systematically fitted together with the $ p $-$ h $ part of the EDF
until Reinhard and Nazarewicz produced a set of new zero-range pairing interactions,
obtained with different fitting procedures but keeping the same EDF form as that of {\FaNDF}~\cite{
  Reinhard2017Phys.Rev.C95_064328}.
Such EDFs were used to interpret the charge radii observed in isotopic chains by laser spectroscopy in a series of papers~\cite{
  Hammen2018Phys.Rev.Lett.121_102501,
  Miller2019Nat.Phys.15_432,
  Gorges2019Phys.Rev.Lett.122_192502,
  Koszorus2021Nat.Phys.17_439,
  Geldhof2022Phys.Rev.Lett.128_152501,
  Reponen2021Nat.Commun.12_4596,
  Hur2022Phys.Rev.Lett.128_163201,
  Sommer2022Phys.Rev.Lett.129_132501,
  Malbrunot-Ettenauer2022Phys.Rev.Lett.128_022502,
  Koenig2023Phys.Rev.Lett.131_102501,
  Karthein2024Nat.Phys.20_1719}.
It should be remarked that the calculations were carried out in different ways,
using either the Bardeen-Cooper-Schrieffer (BCS) or Hartree-Fock-Bogoliubov (HFB) formalism,
and different cutoff strategies,
where the cutoff strategy is crucial for a convergence problem~\cite{
  Dobaczewski1996Phys.Rev.C53_2809}.
Most recently, the effect of an isospin-dependent term in the pairing interaction~\cite{
  Reinhard2024J.Phys.G51_105101}
and
that of a finite-range pairing interaction~\cite{
  Lalit2026Phys.Rev.C113_054310}
were considered.
\par
The present work presents a detailed analysis of the pairing part of the {\FaNDF} EDF
by studying the dependence of pairing gaps and charge radii along the isotopic chains of $ \mathrm{Ca} $, $ \mathrm{Sn} $, and $ \mathrm{Pb} $ nuclei on the parameters of the pairing interaction,
thus extending the recent study by Inakura \textit{et al.}~\cite{
  Inakura2024Phys.Rev.C110_054315}.
Our aim is not the optimization of the pairing interaction,
but a better understanding of the consequences of its peculiar density dependence,
also studying in detail the rearrangement potential.
The results obtained with the Fayans EDF are intriguing,
in particular because it appears to be the only EDF able to reproduce quantitatively the bell-shaped dependence of the charge radii of $ \mathrm{Ca} $ isotopes between $ \nuc{Ca}{40}{} $ and $ \nuc{Ca}{48}{} $.
It is important to understand the differences between the rearrangement potentials associated with the Fayans EDF,
with respect to those produced by ordinary mixed- or surface-type pairing interactions,
which are not able to reproduce the experimental bell shape.
\par
We remark that there have been several attempts to reproduce the bell shape of $ \mathrm{Ca} $ radii other than those based on DFT.
Calculations based on the inclusion of the effects of zero-point energy fluctuations on the density have been
at least partially successful~\cite{
  Khodel1982J.Phys.G8_967,
  Barranco1985Phys.Lett.B151_90,
  Co2022Phys.Rev.C105_034320,
  Brown2022Phys.Rev.C106_L011304,
  Xie2025Phys.Rev.C112_L021303}.
However, these calculations either are not completely self-consistent or have other approximations so that they are not fully conclusive.
It remains an open question 
whether they can be related, at least qualitatively, with the density-dependent pairing interaction included in Fayans EDF (see the related discussions in Refs.~\cite{
  Fayans1996Phys.Lett.B383_19,
  Fayans2000Nucl.Phys.A676_49}).
\par
A shell-model calculation could reproduce the bell shape (although not as pronounced as in experiment),
which was attributed to the promotion of protons from the $ sd $ shell to the $ pf $ shell~\cite{
  Caurier2001Phys.Lett.B522_240,
  Poves}.
On the other hand, recent in-medium similarity renormalization group calculations~\cite{
  Heinz2025Phys.Rev.C111_034311}
included such excitations but did not reproduce the parabolic shape,
and in general \textit{ab initio} calculations do not reproduce $ \mathrm{Ca} $ charge radii accurately,
also for $ \mathrm{Ca} $ nuclei beyond $ \nuc{Ca}{48}{} $~\cite{
  GarciaRuiz2016Nat.Phys.12_594,
  Miyagi2020Phys.Rev.C102_034320,
  Miyagi2025Front.Phys.13_1581854},
nor do higher-order correlations describe the bell shape.
\par
In conclusion, there is no consensus yet about the origin of the bell shape in the nuclear physics community.
The main aim of the present work is to study the pairing part of the Fayans EDF
and to explore whether it is possible to find a parametrization that provides
a good reproduction of the charge radii along the $ \mathrm{Ca} $, $ \mathrm{Sn} $, and $ \mathrm{Pb} $ isotopic chains,
reproducing simultaneously the empirical pairing gaps deduced from the experimental odd-even mass difference.
In practice, we explore a large variety of parameter sets,
and conclude that none of them can reproduce the charge radii of these three isotopic chains simultaneously.
Furthermore, we focus on the rearrangement effect originating from the pairing channel of EDF,
in order to clarify the physical mechanism behind the bell-shaped structure of the charge radii of $ \mathrm{Ca} $ isotopes.
We try to investigate whether this peculiar effect can be mimicked by a refit of the pairing strength.
Throughout the paper, we use the original Fayans EDF named {\FaNDF}~\cite{
  Fayans1998JETPLett.68_169}
in the $ p $-$ h $ channel. 
\par
This paper is organized as follows:
The theoretical framework will be given in Sec.~\ref{sec:theoretical}.
In Sec.~\ref{sec:results}, the calculation results will be given:
The pairing strength will be determined in Sec.~\ref{sec:strength};
the charge radii will be discussed in Sec.~\ref{sec:charge_radii};
then the role of the rearrangement potential and the mechanism of the Fayans $ p $-$ p $ EDF to describe the bell shape will be discussed in Sec.~\ref{sec:rea}.
Finally, in Sec.~\ref{sec:summary}, we summarize our results.
%
%
\section{Theoretical Framework}
\label{sec:theoretical}
\subsection{Fayans energy density functional}
\label{sec:Fayans}
\par
The local-type Fayans EDF~\cite{
  Fayans1998JETPLett.68_169}
can be written as
\begin{equation}
  \label{eq:EDF_all}
  \ca{E}_{\urm{tot}} \left( \ve{r} \right)
  = 
  \ca{E}_{\urm{kin}} \left( \ve{r} \right)
  +
  \ca{E}_{\urm{$ p $-$ h $}} \left( \ve{r} \right)
  +
  \ca{E}_{\urm{$ p $-$ p $}} \left( \ve{r} \right),
\end{equation}
where $ \ca{E}_{\urm{kin}} $, $ \ca{E}_{\urm{$ p $-$ h $}} $ and $ \ca{E}_{\urm{$ p $-$ p $}} $
denote the contributions from the kinetic energy,
the $ p $-$ h $ channel,
and
the $ p $-$ p $ channel, respectively.
The kinetic-energy term $ \ca{E}_{\urm{kin}} $ reads
\begin{equation}
  \label{eq:EDF_kin}
  \ca{E}_{\urm{kin}} \left( \ve{r} \right)
  =
  \frac{\hbar^2}{2m}
  \left[
    \tau_p \left( \ve{r} \right)
    +
    \tau_n \left( \ve{r} \right)
  \right],
\end{equation}
where $ \tau_p $ and $ \tau_n $ are, respectively, the proton and neutron kinetic densities.
The densities appearing in this paper are defined, for instance, in Refs.~\cite{
  Dobaczewski1984Nucl.Phys.A422_103,
  Chabanat1998Nucl.Phys.A635_231,
  Bender2003Rev.Mod.Phys.75_121}.
Note that the center-of-mass (c.m.) correction is not considered.
\subsubsection{Particle-hole channel}
\label{sec:Fayans_ph}
\par
The Fayans $ p $-$ h $ channel is divided into three terms:
\begin{widetext}
  \begin{equation}
    \label{eq:EDF_ph}
    \ca{E}_{\urm{$ p $-$ h $}} \left( \ve{r} \right)
    =
    \ca{E}_{\urm{cent}}
    \left( \alpha \left( \ve{r} \right), \nabla \alpha \left( \ve{r} \right),
      \beta \left( \ve{r} \right), \nabla \beta \left( \ve{r} \right) \right)
    +
    \ca{E}_{\urm{LS}}
    \left( \rho \left( \ve{r} \right), \rho_{\urm{IV}} \left( \ve{r} \right),
      \ve{J} \left( \ve{r} \right), \ve{J}_{\urm{IV}} \left( \ve{r} \right) \right)
    +
    \ca{E}_{\urm{Coul}}
    \left( \rho_p \left( \ve{r} \right), \rho_n \left( \ve{r} \right) \right).
  \end{equation}
  The central part $ \ca{E}_{\urm{cent}} $ is further divided into two parts:
  the volume and surface terms.
  The volume term is fitted to a microscopic calculation of the nuclear equation of state.
  The EDF form of the surface term is determined by the Pad\'{e} approximation,
  and the parameters are fitted to experimental data.
  The resulting central part $ \ca{E}_{\urm{cent}} $ and the spin-orbit one $ \ca{E}_{\urm{LS}} $ are, respectively,
  \begin{subequations}
    \begin{align}
      \ca{E}_{\urm{cent}} \left( \alpha, \nabla \alpha, \beta, \nabla \beta \right)
      & =
        \frac{1}{3}
        \epsilon_{\urm{F}}
        \rho_0
        \left[
        a_{\alpha}^{\urm{V}}
        f_{\alpha} \left( \alpha \right)
        \alpha^2
        +
        a_{\beta}^{\urm{V}}
        f_{\beta} \left( \alpha \right)
        \beta^2
        +
        a_{\alpha}^{\urm{S}}
        r_{\urm{s}}^2
        g_{\alpha} \left( \alpha, \nabla \alpha \right)
        \left| \nabla \alpha \right|^2 
        +
        a_{\beta}^{\urm{S}}
        r_{\urm{s}}^2
        g_{\beta} \left( \alpha, \nabla \alpha \right)
        \left| \nabla \beta \right|^2 
        \right],
        \label{eq:Fayans_cent} \\
      \ca{E}_{\urm{LS}} \left( \rho, \rho_{\urm{IV}}, \ve{J}, \ve{J}_{\urm{IV}} \right)
      & =
        \frac{4 \epsilon_{\urm{F}} r_{\urm{s}}^2}{3 \rho_0}
        \left[
        \kappa_0
        \rho
        \nabla \cdot \ve{J}
        +
        \kappa_1
        \rho_{\urm{IV}}
        \nabla \cdot \ve{J}_{\urm{IV}}
        +
        g_0
        \ve{J}^2
        +
        g_1
        \ve{J}_{\urm{IV}}^2
        \right]
        \label{eq:Fayans_ls}
    \end{align}
  \end{subequations}
\end{widetext}
with
\begin{align}
  f_j \left( \alpha \right)
  & =
    \frac{1 - b_j^{\urm{V}} \alpha^{\sigma_j^{\uurm{V}}}}
    {1 + c_j^{\urm{V}} \alpha^{\sigma_j^{\uurm{V}}}}, \\ 
  g_j \left( \alpha, \nabla \alpha \right)
  & =
    \frac{1}{1 + c_j^{\urm{S}} \alpha^{\sigma_j^{\uurm{S}}}
    + d_j^{\urm{S}} r_{\urm{s}}^2 \left| \nabla \alpha \right|^2}
\end{align}
($ j = \alpha $, $ \beta $).
Here, $ \rho_0 = 0.16 \, \mathrm{fm}^{-3} $ is used
and 
$ r_{\urm{s}} $ and $ \epsilon_{\urm{F}} $ are the Wigner-Seitz radius and the Fermi energy, respectively,
defined by
\begin{align}
  r_{\urm{s}}
  & =
    \left(
    \frac{3}{4 \pi \rho_0}
    \right)^{1/3}, \\
  \epsilon_{\urm{F}}
  & =
    \left(
    \frac{9 \pi}{8}
    \right)^{2/3}
    \frac{\hbar^2}{2m r_{\urm{s}}^2}.
\end{align}
The normalized densities $ \alpha $ and $ \beta $ are, respectively, defined by
\begin{align}
  \alpha
  & =
    \frac{\rho_n + \rho_p}{\rho_0}, \\
  \beta
  & =
    \frac{\rho_n - \rho_p}{\rho_0}
\end{align}
with proton and neutron particle densities $ \rho_p $ and $ \rho_n $.
The parameters to be determined are
$ a_j^{\urm{V}} $, $ b_j^{\urm{V}} $, $ c_j^{\urm{V}} $, $ \sigma_j^{\urm{V}} $, 
$ a_j^{\urm{S}} $, $ c_j^{\urm{S}} $, $ d_j^{\urm{S}} $, $ \sigma_j^{\urm{S}} $
($ j = \alpha $, $ \beta $)
for the central part
and
$ \kappa_j $ and $ g_j $ ($ j = 0 $, $ 1 $) for the spin-orbit part.
For the sake of generality, we introduce the isovector surface term [the $ a_{\beta}^{\urm{S}} $ term in Eq.~\eqref{eq:Fayans_cent}]
and the $ \ve{J}^2 $ terms [the third and fourth terms of Eq.~\eqref{eq:Fayans_ls}],
while none of the proposed Fayans EDFs~\cite{
  Fayans1998JETPLett.68_169,
  Reinhard2017Phys.Rev.C95_064328,
  Reinhard2024J.Phys.G51_105101}
have considered them.
All the proposed Fayans EDFs assumed $ \sigma_{\alpha}^{\urm{S}} = 1/3 $.
The original Fayans EDF, {\FaNDF}~\cite{
  Fayans1998JETPLett.68_169},
also fixed $ \kappa_1 = 0 $.
\par
The Coulomb term $ \ca{E}_{\urm{Coul}} $ consists of
the Coulomb Hartree term $ \ca{E}_{\urm{CH}} $ and
the Coulomb exchange term $ \ca{E}_{\urm{Cx}} $
\begin{equation}
  \label{eq:EDF_Coul}
  \ca{E}_{\urm{Coul}}
  \left( \rho_p, \rho_n \right)
  = 
  \ca{E}_{\urm{CH}}
  \left( \rho_p \right)
  +
  \ca{E}_{\urm{Cx}}
  \left( \rho_p, \alpha \right).
\end{equation}
The proton density is used for $ \ca{E}_{\urm{CH}} $ as 
\begin{equation}
  \label{eq:EDF_CH}
  \ca{E}_{\urm{CH}}
  \left( \rho_p \left( \ve{r} \right) \right)
  =
  \frac{e^2}{2}
  \int
  \frac{\rho_p \left( \ve{r} \right) \rho_p \left( \ve{r}' \right)}{\left| \ve{r} - \ve{r}' \right|}
  \, d \ve{r}'.
\end{equation}
While the exact Coulomb exchange is nonlocal and spoils the simplicity of quasilocal EDFs,
in most cases the Coulomb exchange energy is included by adopting the Slater approximation;
nevertheless, the Coulomb exchange term of the {\FaNDF}~\cite{
  Fayans1998JETPLett.68_169}
reads
\begin{equation}
  \label{eq:EDF_Cx}
  \ca{E}_{\urm{Cx}}
  \left( \rho_p, \alpha \right)
  =
  - \frac{3 e^2}{4}
  \left(
    \frac{3}{\pi}
  \right)^{1/3}
  \rho_p^{4/3}
  \left(
    1 - h_{\urm{C}} \alpha^{\sigma_{\uurm{C}}}
  \right),
\end{equation}
implicitly including a Coulomb-nuclear correlation term~\cite{
  Fayans1998JETPLett.68_169}
via the $ h_{\urm{C}} $ term~\cite{
  Bulgac1996Nucl.Phys.A601_103,
  Bulgac1999Phys.Lett.B469_1,
  Reinhard2017Phys.Rev.C95_064328},
which was also claimed to mock up effectively isospin-symmetry-breaking effects and reproduce the Okamoto-Nolen-Schiffer anomaly in some mirror nuclei.
Here, $ h_{\urm{C}} $ and $ \sigma_{\urm{C}} $ are parameters to be fitted,
and if $ h_{\urm{C}} $ is set equal to zero, one recovers the usual Hartree-Fock-Slater approximation.
\subsubsection{Particle-particle channel}
\label{sec:Fayans_pp}
\par
The Fayans EDF in the $ p $-$ p $ channel is originally defined by
\begin{align}
  & \ca{E}_{\urm{$ p $-$ p $}} 
    \left( \alpha, \nabla \alpha, \tilde{\rho}_p, \tilde{\rho}_n \right)
    \notag \\
  & =
    \frac{\epsilon_{\urm{F}}}{3 \rho_0}
    \left(
    f
    +
    h_0
    \alpha^{\gamma}
    +
    h_{\urm{D}}
    r_{\urm{s}}^2
    \left| \nabla \alpha \right|^2
    \right)
    \sum_q
    \tilde{\rho}_q^2
    \notag \\
  & =
    \frac{\epsilon_{\urm{F}} f}{3 \rho_0}
    \left(
    1
    -
    \tilde{h}_0
    \alpha^{\gamma}
    -
    \tilde{h}_{\urm{D}}
    r_{\urm{s}}^2
    \left| \nabla \alpha \right|^2
    \right)
    \sum_q
    \tilde{\rho}_q^2,
    \label{eq:EDF_pair}
\end{align}
where $ \tilde{\rho}_p $ and $ \tilde{\rho}_n $ are, respectively, the proton and neutron pairing densities.
The parameter $ f < 0 $ determines the overall strength of the pairing interaction,
while the parameters $ \tilde{h}_0 = - h_0/f $ and $ \tilde{h}_{\urm{D}} = - h_{\urm{D}}/f $ determine its volume and/or surface character.
\par
The widely-used density-dependent delta pairing interaction~\cite{
  Dobaczewski1984Nucl.Phys.A422_103}
is
\begin{equation}
  v_{\urm{$ p $-$ p $}} \left( \ve{r} \right) 
  =
  V_0
  \left(
    1
    -
    \eta
    \frac{\rho}{\rho_0}
  \right)
  \delta \left( \ve{r} \right) ;
\end{equation}
accordingly, the EDF $ \ca{E}_{\urm{$ p $-$ p $}} $ reads
\begin{equation}
  \ca{E}_{\urm{$ p $-$ p $}} 
  \left( \rho, \tilde{\rho}_p, \tilde{\rho}_n \right)
  =
  \frac{V_0}{4}
  \left(
    1
    -
    \eta
    \frac{\rho}{\rho_0}
  \right)
  \sum_q
  \tilde{\rho}_q^2,
\end{equation}
where $ \eta = 0 $, $ 1/2 $, and $ 1 $ correspond to
the volume-, mixed-, and surface-type pairings, respectively.
Therefore, Eq.~\eqref{eq:EDF_pair} with $ \gamma = 1 $ and $ \tilde{h}_{\urm{D}} = 0 $ corresponds to the standard density-dependent delta pairing interaction,
where $ \tilde{h}_0 = 0 $, $ 1/2 $, and $ 1 $ are, respectively,
volume-, mixed-, and surface-type pairing interactions.
\par
We introduce different strengths for protons and for neutrons
to account for the Coulomb antipairing effect effectively~\cite{
  Anguiano2001Nucl.Phys.A683_227,
  Nakada2011Phys.Rev.C83_031302},
while keeping $ \tilde{h}_0 $ and $ \tilde{h}_{\urm{D}} $ the same for the two channels.
Namely,
we use the different strengths
$ f_q = h_q f $ ($ q = p $, $ n $)
with the parameter $ h_p \in \left( 0, 1 \right] $, i.e., $ \left| f_p \right| \le \left| f_n \right| $,
to consider the Coulomb antipairing
($ h_n = 1 $ is always kept).
Then, the pairing EDF we use in this paper is 
\begin{align}
  & \ca{E}_{\urm{$ p $-$ p $}} 
    \left( \alpha, \nabla \alpha, \tilde{\rho}_p, \tilde{\rho}_n \right)
    \notag \\
  & =
    \frac{\epsilon_{\urm{F}}}{3 \rho_0}
    \left(
    1
    -
    \tilde{h}_0
    \alpha^{\gamma}
    -
    \tilde{h}_{\urm{D}}
    r_{\urm{s}}^2
    \left| \nabla \alpha \right|^2
    \right)
    \sum_q
    f_q 
    \tilde{\rho}_q^2
    \notag \\
  & =
    \frac{\epsilon_{\urm{F}} f}{3 \rho_0}
    \left(
    1
    -
    \tilde{h}_0
    \alpha^{\gamma}
    -
    \tilde{h}_{\urm{D}}
    r_{\urm{s}}^2
    \left| \nabla \alpha \right|^2
    \right)
    \sum_q
    h_q
    \tilde{\rho}_q^2.
    \label{eq:EDF_pair2}
\end{align}
In this paper, $ h_p = 0.85 $ is used
and $ f = f_n $ is determined in Sec.~\ref{sec:strength}.
These values are consistent with theoretical studies~\cite{
  Nakada2011Phys.Rev.C83_031302},
which found a Coulomb antipairing effect.
On the other hand, previous fits made with functionals different from Fayans pointed out 
that the proton pairing strength should be larger than the neutron one~\cite{
  Bertsch2009Phys.Rev.C79_034306,
  Kluepfel2009Phys.Rev.C79_034310}.
However, we have found that using larger values of $ h_p $ may lead to finite values of the proton gap
in $ \mathrm{Ca} $ isotopes for some of the adopted pairing interactions,
including surface pairing ($ \tilde{h}_0 = 1.00 $, $ \tilde{h}_{\urm{D}} = 0.00 $).
\par
The functional derivative of $ \ca{E}_{\urm{$ p $-$ p $}} $ yields the mean-field potential.
That with respect to the $ \tilde{\rho}_q $ ($ q = p $, $ n $) is the pairing mean-field potential
\begin{equation}
  \label{eq:pot_pp}
  \tilde{V}_q \left( \ve{r} \right)
  =
  \frac{2 \epsilon_{\urm{F}} h_q f}{3 \rho_0}
  \left\{
    1
    -
    \tilde{h}_0
    \left[ \alpha \left( \ve{r} \right) \right]^{\gamma}
    -
    \tilde{h}_{\urm{D}}
    r_{\urm{s}}^2
    \left| \nabla \alpha \left( \ve{r} \right) \right|^2
  \right\}
  \tilde{\rho}_q \left( \ve{r} \right).
\end{equation}
Because of the existence of the density-dependent terms ($ \tilde{h}_0 $ and $ \tilde{h}_{\urm{D}} $ terms)
in Eq.~\eqref{eq:EDF_pair2},
$ \ca{E}_{\urm{$ p $-$ p $}} $ also contributes to the $ p $-$ h $ mean-field potential,
\begin{widetext}
  \begin{align}
    V_q^{\urm{rea}} \left( \ve{r} \right)
    & =
      -
      \frac{\epsilon_{\urm{F}}}{3 \rho_0}
      \left(
      \frac{\gamma \tilde{h}_0}{\rho_0} 
      \alpha^{\gamma - 1}
      \sum_{q'}
      f_{q'}
      \left[
      \tilde{\rho}_{q'} \left( \ve{r} \right)
      \right]^2
      -
      \frac{2 \tilde{h}_{\urm{D}}}{\rho_0} 
      r_{\urm{s}}^2
      \nabla
      \cdot
      \left\{
      \nabla \alpha
      \cdot
      \sum_{q'}
      f_{q'}
      \left[
      \tilde{\rho}_{q'} \left( \ve{r} \right)
      \right]^2
      \right\}
      \right)
      \notag \\
    & =
      -
      \frac{\epsilon_{\urm{F}}}{3 \rho_0}
      \left(
      \frac{\gamma \tilde{h}_0}{\rho_0} 
      \alpha^{\gamma - 1}
      \sum_{q'}
      f_{q'}
      \left[
      \tilde{\rho}_{q'} \left( \ve{r} \right)
      \right]^2
      -
      \frac{2 \tilde{h}_{\urm{D}}}{\rho_0} 
      r_{\urm{s}}^2
      \left\{
      \laplace \alpha
      \sum_{q'}
      f_{q'}
      \left[
      \tilde{\rho}_{q'} \left( \ve{r} \right)
      \right]^2
      +
      \nabla \alpha
      \cdot
      \nabla
      \sum_{q'}
      f_{q'}
      \left[
      \tilde{\rho}_{q'} \left( \ve{r} \right)
      \right]^2
      \right\}
      \right)
      \notag \\
    & =
      -
      \frac{\epsilon_{\urm{F}} f}{3 \rho_0^2}
      \left\{
      \left(
      \tilde{h}_0 \gamma 
      \alpha^{\gamma - 1}
      -
      2 \tilde{h}_{\urm{D}}
      r_{\urm{s}}^2
      \laplace \alpha
      \right)
      \sum_{q'}
      h_{q'}
      \left[
      \tilde{\rho}_{q'} \left( \ve{r} \right)
      \right]^2
      -
      4 \tilde{h}_{\urm{D}}
      r_{\urm{s}}^2
      \nabla \alpha
      \cdot
      \sum_{q'}
      h_{q'}
      \tilde{\rho}_{q'} \left( \ve{r} \right)
      \nabla
      \tilde{\rho}_{q'} \left( \ve{r} \right)
      \right\},
      \label{eq:pot_rea}
  \end{align}
\end{widetext}
which we call the rearrangement potential.
This rearrangement potential is identical for both protons and neutrons,
$ V^{\urm{rea}} \left( \ve{r} \right) = V^{\urm{rea}}_p \left( \ve{r} \right) = V^{\urm{rea}}_n \left( \ve{r} \right) $,
and
is zero for $ \tilde{h}_0 = \tilde{h}_{\urm{D}} = 0 $, i.e., the volume-type pairing.
\subsubsection{EDF used in this paper}
\label{sec:Fayans_used}
\par
In this paper, we use the original Fayans EDF named {\FaNDF}~\cite{
  Fayans1998JETPLett.68_169}
in the $ p $-$ h $ channel.
The values of the parameters are listed in Table~\ref{tab:FaNDF0},
where the parameters with the value of $ 0.0 $ are not considered even in the fitting procedure.
In the Appendix, we show results obtained with another parametrization of the Fayans EDF named
Fy ($ \Delta r $)~\cite{
  Reinhard2017Phys.Rev.C95_064328}
and with the Skyrme interaction SkM*~\cite{
  Bartel1982Nucl.Phys.A386_79}.
Note that the {\FaNDF} does not consider the c.m.~correction to Eq.~\eqref{eq:EDF_kin},
while the two EDFs used in the Appendix take into account the one-body c.m.~correction in a self-consistent manner.
\par
In Ref.~\cite{
  Fayans1998JETPLett.68_169},
the {\FaNDF} EDF was used with the parameters
$ f_p = f_n = -2.8 $, $ h_0 = 2.8 $, and $ h_{\urm{D}} = 2.2 $,
i.e., $ \tilde{h}_0 = 1.0 $ and $ \tilde{h}_{\urm{D}} = 0.79 $,
with $ \gamma = 1 $.
It should be noted that these values were determined from a previous fit on neutron separation energies and charge radii of $ \mathrm{Pb} $ isotopes~\cite{
  Fayans1996Phys.Lett.B383_19}
and were not fitted consistently with the $ p $-$ h $ part of the EDF.
In the following, we study the results obtained by varying the parameters $ \tilde{h}_0 $ and $ \tilde{h}_{\urm{D}} $ on a mesh in the interval between zero and one,
using the values $ 0 $, $ 1/4 $, $ 1/2 $, $ 3/4 $, and $ 1 $,
for a total of $ 25 $ parameter sets.
The parameter $ \gamma $ is always fixed to one.
For each parameter set, the overall strength $ f $ will be adjusted to reproduce the experimental neutron pairing gap of $ \nuc{Ca}{44}{} $.
The details of this fitting procedure are explained below.
\begin{table}[tb]
  \centering
  \caption{Particle-hole parameters of {\FaNDF} EDF~\cite{
      Fayans1998JETPLett.68_169}.}
  \label{tab:FaNDF0}
  \begin{ruledtabular}
    \begin{tabular}{cd}
      \multicolumn{1}{l}{Parameter} & \multicolumn{1}{l}{Value} \\
      \hline
      $ a_{\alpha}^{\urm{V}} $      & -9.559  \\
      $ b_{\alpha}^{\urm{V}} $      &  0.633  \\
      $ c_{\alpha}^{\urm{V}} $      &  0.131  \\
      $ \sigma_{\alpha}^{\urm{V}} $ & 1/3     \\
      \hline
      $ a_{\beta}^{\urm{V}} $       &  4.428  \\
      $ b_{\beta}^{\urm{V}} $       &  0.250  \\
      $ c_{\beta}^{\urm{V}} $       &  1.300  \\
      $ \sigma_{\beta}^{\urm{V}} $  &  1.0    \\
      \hline
      $ a_{\alpha}^{\urm{S}} $      &  0.600  \\
      $ c_{\alpha}^{\urm{S}} $      &  0.131  \\
      $ d_{\alpha}^{\urm{S}} $      &  0.440  \\
      $ \sigma_{\alpha}^{\urm{S}} $ & 1/3     \\
      \hline
      $ a_{\beta}^{\urm{S}} $       &  0.0    \\
      $ c_{\beta}^{\urm{S}} $       &  0.0    \\
      $ d_{\beta}^{\urm{S}} $       &  0.0    \\
      $ \sigma_{\beta}^{\urm{S}} $  &  0.0    \\
      \hline
      $ \kappa_0 $                  & -0.19   \\
      $ \kappa_1 $                  &  0.0    \\
      $ g_0 $                       &  0.0    \\
      $ g_1 $                       &  0.0    \\
      \hline
      $ h_{\urm{C}} $               &  0.941  \\
      $ \sigma_{\urm{C}} $          & 1/3     \\
    \end{tabular}
  \end{ruledtabular}
\end{table}
\subsection{Numerical setup}
\label{sec:numerical}
\par
The Hartree-Fock-Bogoliubov (HFB) equation associated with the {\FaNDF} EDF
is solved assuming spherical symmetry
in a $ 16 \, \mathrm{fm} $ box on a radial mesh with a step equal to $ 0.1 \, \mathrm{fm} $.
The HFB equation is solved directly via the diagonalization of the HFB matrix.
Single-particle states up to $ E_{\urm{cut}} = 60 \, \mathrm{MeV} $ in the Hartree-Fock equivalent energy $ \epsilon_j $ are taken into account for the calculation~\cite{
  Stoitsov2013Comput.Phys.Commun.184_1592}.
We have carefully checked the dependence on $ E_{\urm{cut}} $,
finding that our results for quasiparticle energies and pairing gaps have converged,
so that a detailed analysis of this point will not be required in the following.
\par
Note that, throughout this paper, the normal and pairing densities
under the spherical symmetry 
are, respectively, defined by
\begin{subequations}
  \begin{align}
    \rho_q \left( r \right)
    & =
      \frac{1}{r^2}
      \sum_{\epsilon_j < E_{\uurm{cut}}}
      \left( 2j + 1 \right)
      \left[ v_j \left( r \right) \right]^2, \\
    \tilde{\rho}_q \left( r \right)
    & =
      -
      \frac{1}{r^2}
      \sum_{\epsilon_j < E_{\uurm{cut}}}
      \left( 2j + 1 \right)
      u_j \left( r \right)
      v_j \left( r \right),
  \end{align}
\end{subequations}
where only the states with the positive quasiparticle energy are considered in the summation.
Here,
$ u_j $ and $ v_j $ are, respectively, the upper and lower components of an HFB eigenfunction
with the normalization
\begin{equation}
  4 \pi 
  \int_0^{\infty}
  \left\{
    \left[ v_j \left( r \right) \right]^2
    +
    \left[ u_j \left( r \right) \right]^2
  \right\}
  \, dr
  =
  1.
\end{equation}
%
%
%
\section{Calculation Results}
\label{sec:results}
\par
Before going to the calculation,
we define the relative difference of mean-square charge radius with respect to a magic nucleus
($ \nuc{Ca}{40}{} $, $ \nuc{Sn}{132}{} $, or $ \nuc{Pb}{208}{} $) as
\begin{subequations}
  \begin{align}
    \delta^2 \left( \nuc{Ca}{A}{} \right)
    & =
      R_{\urm{ch}}^2 \left( \nuc{Ca}{A}{} \right) - R_{\urm{ch}}^2 \left( \nuc{Ca}{40}{} \right), \\
    \delta^2 \left( \nuc{Sn}{A}{} \right)
    & =
      R_{\urm{ch}}^2 \left( \nuc{Sn}{A}{} \right) - R_{\urm{ch}}^2 \left( \nuc{Sn}{132}{} \right), \\
    \delta^2 \left( \nuc{Pb}{A}{} \right)
    & =
      R_{\urm{ch}}^2 \left( \nuc{Pb}{A}{} \right) - R_{\urm{ch}}^2 \left( \nuc{Pb}{208}{} \right).
  \end{align}
\end{subequations}
The charge radii are calculated theoretically via~\cite{  
  Bertozzi1972Phys.Lett.B41_408,
  Chabanat1997Nucl.Phys.A627_710,
  Horowitz2012Phys.Rev.C86_045503,
  Naito2021Phys.Rev.C104_024316}
\begin{equation}
  \label{eq:charge}
  R_{\urm{ch}}^2
  =
  R_p^2
  +
  r_p^2
  +
  \frac{N}{Z}
  r_n^2
  +
  R_{\urm{SO} p}^2
  +
  \frac{N}{Z}
  R_{\urm{SO} n}^2,
\end{equation}
where $ R_p $ is the proton root-mean-square radius,
$ r_p = 0.8409 \, \mathrm{fm} $ is the single-proton root-mean-square charge radius~\cite{
  Navas2024Phys.Rev.D110_030001},
and 
$ r_n^2 = -0.1155 \, \mathrm{fm}^2 $ is the single-neutron mean-square charge radius~\cite{
  Navas2024Phys.Rev.D110_030001}.
The spin-orbit contribution $ R_{\urm{SO} q} $ ($ q = p $, $ n $) is obtained by
\begin{equation}
  R_{\urm{SO} q}^2
  =
  \frac{\kappa_q}{M^2 N_q}
  \sum_i
  \ca{N}_{i q}
  \avr{\ve{l} \cdot \ve{\sigma}}_i 
\end{equation}
with $ N_p = Z $, $ N_n = N $,
the nucleon anomalous magnetic moments $ \kappa_q $
with $ \kappa_p = 1.79284734463 $ and $ \kappa_n = -1.91304273 $~\cite{
  Navas2024Phys.Rev.D110_030001},
and $ \ca{N}_{i q} $ and $ M $ standing for the occupation number of the orbital $ i $ and the nucleon mass, respectively.
\par
As we have previously discussed,
almost all the EDFs used in the literature are not able to reproduce the bell shape of the charge radii between $ \nuc{Ca}{40}{} $ and $ \nuc{Ca}{48}{} $ as a function of the neutron number.
In most cases, the very small experimental difference between the charge radii of the two closed-shell nuclei $ \nuc{Ca}{40}{} $ and $ \nuc{Ca}{48}{} $,
$ \delta^2 \left( \nuc{Ca}{48}{} \right) $,
is overestimated,
as can be observed in Table~\ref{tab:various_EDF}, where the results obtained with various EDFs~\cite{
  Bartel1982Nucl.Phys.A386_79,
  Fayans1998JETPLett.68_169,
  Chabanat1998Nucl.Phys.A635_231,
  Kluepfel2009Phys.Rev.C79_034310,
  Kortelainen2010Phys.Rev.C82_024313,
  Kortelainen2012Phys.Rev.C85_024304,
  Roca-Maza2012Phys.Rev.C86_031306,
  Kortelainen2014Phys.Rev.C89_054314,
  Reinhard2017Phys.Rev.C95_064328}
are reported.
The Fayans EDFs {\FaNDF}, Fy ($ \Delta r $, $ \Delta r^{\urm{oe}} $), and Fy ($ \Delta r $), as well as the Skyrme SkM*, clearly stand out for
their ability to reproduce the experimental value of $ \delta^2 \left( \nuc{Ca}{48}{} \right) $
showing substantially better agreement than most of the EDFs.
This represents a remarkable feature of the $ p $-$ h $ part of these EDFs,
and in the following we investigate the performance of {\FaNDF} in the case of open-shell nuclei.
The case of Fy ($ \Delta r $) will be studied in the Appendix,
where we also consider the SkM* EDF, which also reproduces rather well the experimental result of $ \delta^2 \left( \nuc{Ca}{48}{} \right) $.
\begin{table*}[tb]
  \centering
  \caption{Charge radii $ R_{\urm{ch}} $ of $ \nuc{Ca}{40}{} $ and $ \nuc{Ca}{48}{} $
    and the difference $ \delta^2 \left( \nuc{Ca}{48}{} \right) $
    between their squared charge radii
    calculated by various energy density functionals (EDFs)~\cite{
      Bartel1982Nucl.Phys.A386_79,
      Fayans1998JETPLett.68_169,
      Chabanat1998Nucl.Phys.A635_231,
      Kluepfel2009Phys.Rev.C79_034310,
      Kortelainen2010Phys.Rev.C82_024313,
      Kortelainen2012Phys.Rev.C85_024304,
      Roca-Maza2012Phys.Rev.C86_031306,
      Kortelainen2014Phys.Rev.C89_054314,
      Reinhard2017Phys.Rev.C95_064328}.
    For comparison, the experimental data~\cite{
      Angeli2013At.DataNucl.DataTables99_69}
    are also shown.}
  \label{tab:various_EDF}
  \begin{ruledtabular}
    \begin{tabular}{lddd}
      \multicolumn{1}{l}{EDF} & \multicolumn{1}{c}{$ R_{\urm{ch}} \left( \nuc{Ca}{40}{} \right) $ ($ \mathrm{fm} $)} & \multicolumn{1}{c}{$ R_{\urm{ch}} \left( \nuc{Ca}{48}{} \right) $ ($ \mathrm{fm} $)} & \multicolumn{1}{c}{$ \delta^2 \left( \nuc{Ca}{48}{} \right) $ ($ \mathrm{fm}^2 $)} \\
      \hline
      UNEDF0                                     & 3.5104 & 3.5053 & -0.0355 \\
      Fy ($ \Delta r $, $ \Delta r^{\urm{oe}} $) & 3.4604 & 3.4614 & +0.0068 \\
      {\FaNDF}                                   & 3.4843 & 3.4864 & +0.0143 \\
      Fy ($ \Delta r $)                          & 3.4428 & 3.4450 & +0.0148 \\
      SkM*                                       & 3.5115 & 3.5146 & +0.0216 \\
      SV-min                                     & 3.5028 & 3.5129 & +0.0710 \\
      SV-bas                                     & 3.4950 & 3.5081 & +0.0915 \\
      SLy4                                       & 3.5052 & 3.5220 & +0.1186 \\
      SLy5                                       & 3.5013 & 3.5199 & +0.1306 \\
      UNEDF1                                     & 3.4837 & 3.5037 & +0.1399 \\
      Fy (std)                                   & 3.4702 & 3.4921 & +0.1528 \\
      UNEDF2                                     & 3.4736 & 3.4992 & +0.1787 \\
      SAMi                                       & 3.4758 & 3.5054 & +0.2070 \\
      \hline                                  
      Expt.                                      & 3.4776 & 3.4771 & -0.0035 \\ 
    \end{tabular}
  \end{ruledtabular}
\end{table*}
\subsection{Pairing gap}
\label{sec:strength}
\par
We start by calculating the neutron pairing gaps of the three open-shell nuclei representative
of different mass regions:
$ \nuc{Ca}{44}{} $, $ \nuc{Sn}{120}{} $, and $ \nuc{Pb}{204}{}$.
We use the average gap 
\begin{equation}
  \label{eq:average_gap}
  \Delta_q
  =
  \left|
    \frac{\int \rho_q \left( \ve{r} \right) \tilde{V}_q \left( \ve{r} \right) \, d \ve{r}}{\int \rho_q \left( \ve{r} \right) \, d \ve{r}}
  \right|
\end{equation}
to calculate the pairing gap theoretically ($ q = p $, $ n $).
\subsubsection{Determination of the overall strength}
\label{sec:gap_overall}
\par
The theoretical values obtained with {\FaNDF} and the 25 different pairing parameter sets
will be compared with the odd-even shifted three-point mass formula based on
the experimental binding energy $ B \left( Z, N \right) $~\cite{
  Huang2021Chin.Phys.C45_030002,
  Wang2021Chin.Phys.C45_030003}
of a nucleus with $ Z $ protons and $ N $ neutrons,
\begin{equation}
  \label{eq:gap}
  \Delta_n \left( Z, N \right)
  =
  \frac{B \left( Z, N + 2 \right) - 2 B \left( Z, N + 1 \right) + B \left( Z, N \right)}{2},
\end{equation}
leading to 
\begin{subequations}
  \label{eq:gap_ref}
  \begin{align}
    \Delta_n^{\urm{ref}} \left( 20, 24 \right)
    & \simeq
      1.5 \, \mathrm{MeV}, 
      \label{eq:gap_ref_Ca} \\
    \Delta_n^{\urm{ref}} \left( 50, 70 \right)
    & \simeq
      1.3 \, \mathrm{MeV}, 
      \label{eq:gap_ref_Sn} \\
    \Delta_n^{\urm{ref}} \left( 82, 122 \right)
    & \simeq
      0.7 \, \mathrm{MeV}.
      \label{eq:gap_ref_Pb} 
  \end{align}
\end{subequations}
Note that using more accurate values from the experimental masses would not be meaningful,
given the approximate character of the correspondence of the odd-even mass difference with the theoretical gaps
[Eq.~\eqref{eq:average_gap}]~\cite{
  Satula1998Phys.Rev.Lett.81_3599,
  Bender2000Eur.Phys.J.A8_59,
  Duguet2001Phys.Rev.C65_014310,
  Duguet2001Phys.Rev.C65_014311,
  Hilaire2002Phys.Lett.B531_61}.
\par
For each pair of values $ \left( \tilde{h}_0, \tilde{h}_{\urm{D}} \right) $,
we fix the value of the pairing strength $ f $ so that the theoretical gap in $ \nuc{Ca}{44}{} $ reproduces the value of
$ \Delta_n^{\urm{ref}} \left( 20, 24 \right) \simeq 1.5 \, \mathrm{MeV} $.\footnote{
  There are several ways to determine the pairing strength,
  and in this paper we use $ \Delta_n $ of $ \nuc{Ca}{44}{} $.
  This is because one of the main motivations of our study is to investigate the bell shape of the charge radii of $ \mathrm{Ca} $ isotopes,
  and, thus, we would like to make sure that all the pairing interactions used reproduce, at least, one of the quantities of $ \nuc{Ca}{44}{} $.}
The resulting values are summarized in Table~\ref{tab:strength_FaNDF0}.
One can find that the larger $ \tilde{h}_0 $ and $ \tilde{h}_{\urm{D}} $ require larger $ \left| f \right| $.
This is because
the neutron pairing gap [Eq.~\eqref{eq:average_gap}] can be written as
\begin{widetext}
  \begin{align}
    \Delta_n
    & =
      \frac{1}{24}
      \left|
      \int \rho_n \left( \ve{r} \right) \tilde{V}_n \left( \ve{r} \right) \, d \ve{r}
      \right|
      \notag \\
    & =
      \frac{\epsilon_{\urm{F}} \left| f \right|}{36 \rho_0}
      \left|
      \int
      \left\{
      1
      -
      \tilde{h}_0
      \left[ \alpha \left( \ve{r} \right) \right]^{\gamma}
      -
      \tilde{h}_{\urm{D}}
      r_{\urm{s}}^2
      \left| \nabla \alpha \left( \ve{r} \right) \right|^2
      \right\}
      \rho_n \left( \ve{r} \right)
      \tilde{\rho}_n \left( \ve{r} \right)
      \, d \ve{r}
      \right|.
      \label{eq:pairing_gap}
  \end{align}
\end{widetext}
The dependence of the pairing gap on the pairing parameters is mostly determined
by the term in the curly bracket in Eq.~\eqref{eq:pairing_gap},
although the densities are also modified in the self-consistent calculation.
In addition, $ r_{\urm{s}}^2 \left| \nabla \alpha \left( \ve{r} \right) \right|^2 $
is nonnegative;
accordingly, larger $ \tilde{h}_0 $ and $ \tilde{h}_{\urm{D}} $
give smaller
$ \left\{
  1
  -
  \tilde{h}_0
  \left[ \alpha \left( \ve{r} \right) \right]^{\gamma}
  -
  \tilde{h}_{\urm{D}}
  r_{\urm{s}}^2
  \left| \nabla \alpha \left( \ve{r} \right) \right|^2
\right\} $.
Consequently,
larger values of $ \left| f \right| $ are required to reproduce the same $ \Delta_n $.
We also find that the effect of $ \tilde{h}_0 $ on $ f $ is larger than that of $ \tilde{h}_{\urm{D}} $,
and the $ \tilde{h}_{\urm{D}} $ effect is larger for $ \tilde{h}_0 = 1 $ than that for $ \tilde{h}_0 = 0 $.
\par
\begin{figure*}[tb]
  \centering
  \begin{minipage}{0.49\linewidth}
    \centering
    \includegraphics[width=1.0\linewidth]{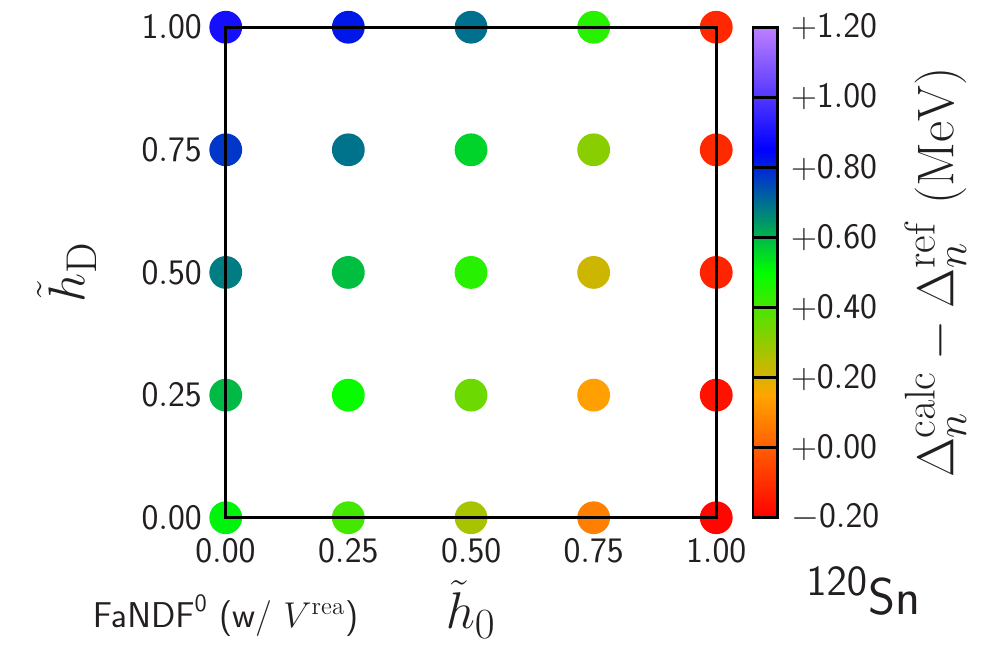}
  \end{minipage}
  \hfill
  \begin{minipage}{0.49\linewidth}
    \centering
    \includegraphics[width=1.0\linewidth]{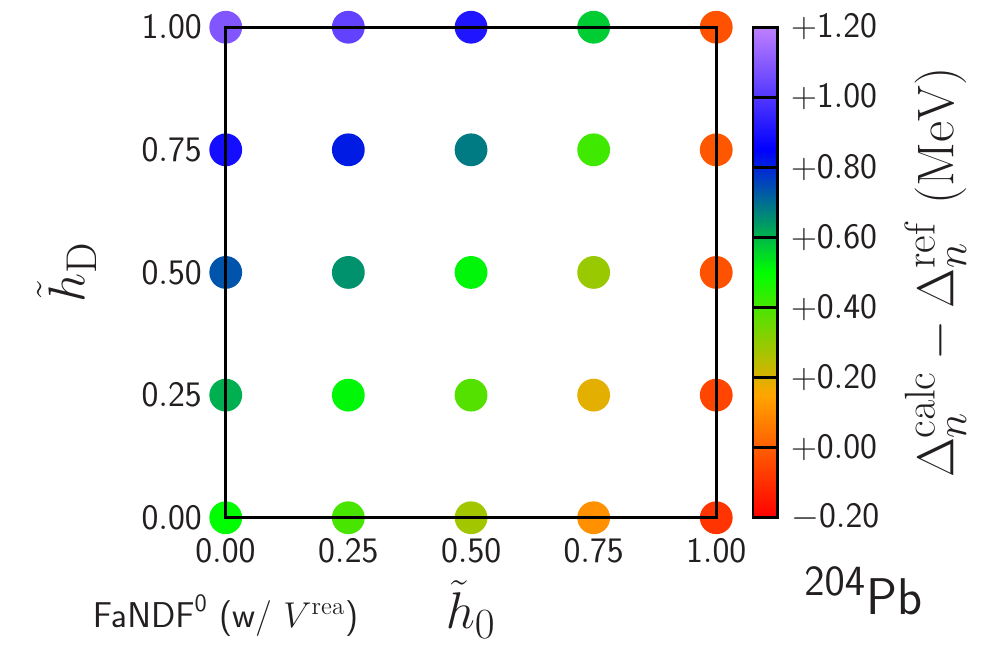}
  \end{minipage}
  \caption{Difference between calculated neutron pairing gap $ \Delta_n^{\urm{calc}} $ and reference one $ \Delta_n^{\urm{ref}} $ of $ \nuc{Sn}{120}{} $ (left) and $ \nuc{Pb}{204}{} $ (right)
    calculated with various pairs of $ \tilde{h}_0 $ and $ \tilde{h}_{\urm{D}} $
    on top of the {\FaNDF} EDF.}
  \label{fig:Deltan_benchmark}
\end{figure*}

We then calculate with the same values of $ f $ the pairing gaps of $ \nuc{Sn}{120}{} $ and $ \nuc{Pb}{204}{} $,
as well as doubly magic nuclei $ \nuc{Ca}{40}{} $, $ \nuc{Sn}{132}{} $, and $ \nuc{Pb}{208}{} $,
which will be used as pivotal nuclei.
The differences between theory ($ \Delta_n^{\urm{calc}} $)
and experiment ($ \Delta_n^{\urm{ref}} $)
obtained with the 25 parameter sets for $ \nuc{Sn}{120}{} $ and $ \nuc{Pb}{204}{} $ are shown in Fig.~\ref{fig:Deltan_benchmark}. 
Smaller values of  $ \tilde{h}_0 $ and larger values of $ \tilde{h}_{\urm{D}} $ tend to produce gaps that are too large.
It is also found that the effect of $ \tilde{h}_0 $ on $ \Delta_n^{\urm{calc}} $ is in the opposite direction with respect to that of $ \tilde{h}_{\urm{D}} $.
\par
Among these 25 parameter sets, we select the pairs satisfying the following conditions based on the calculated pairing gap $ \Delta_q^{\urm{calc}} $ ($ q = p $, $ n $):
\begin{enumerate}
  \renewcommand{\labelenumi}{(\arabic{enumi})}
\item satisfying $ \left| \Delta_n^{\urm{calc}} - \Delta_n^{\urm{ref}} \right| < 250 \, \mathrm{keV} $ in $ \nuc{Sn}{120}{} $ and $ \nuc{Pb}{204}{} $; and 
\item giving $ \Delta_n^{\urm{calc}} < 1 \, \mathrm{keV} $ in $ \nuc{Ca}{40}{} $, $ \nuc{Sn}{132}{} $, and $ \nuc{Pb}{208}{} $.
\end{enumerate}
Such pairs are indicated by a star in Table~\ref{tab:strength_FaNDF0}.
Note that all 25 parameter sets give $ \Delta_p^{\urm{calc}} = 0 $ for $ \nuc{Ca}{40}{} $, $ \nuc{Ca}{44}{} $, $ \nuc{Sn}{120}{} $, $ \nuc{Sn}{132}{} $, $ \nuc{Pb}{204}{} $, and $ \nuc{Pb}{208}{} $.
\begin{table}[tb]
  \centering
  \caption{Pairing strength $ f $ determined to reproduce the neutron pairing gap of $ \nuc{Ca}{44}{} $ [Eq.~\eqref{eq:gap_ref_Ca}] with the {\FaNDF} EDF.
    Among them,
    the satisfied conditions shown in the text are shown with \checkmark.
    The strengths that satisfy all the conditions are shown with the star ($ \star $).}
  \label{tab:strength_FaNDF0}
  \begin{ruledtabular}
    \begin{tabular}{dddcccccl}
      \multicolumn{1}{c}{$ \tilde{h}_0 $} & \multicolumn{1}{c}{$ \tilde{h}_{\urm{D}} $} & \multicolumn{1}{c}{$ -f $} & \multicolumn{1}{c}{$ \nuc{Sn}{120}{} $} & \multicolumn{1}{c}{$ \nuc{Pb}{204}{} $} & \multicolumn{1}{c}{$ \nuc{Ca}{40}{} $} & \multicolumn{1}{c}{$ \nuc{Sn}{132}{} $} & \multicolumn{1}{c}{$ \nuc{Pb}{208}{} $} & \\
      \hline
      0.00 & 0.00 & 0.510 &            &            & \checkmark & \checkmark & \checkmark &  \\
      0.00 & 0.25 & 0.531 &            &            & \checkmark & \checkmark & \checkmark &  \\
      0.00 & 0.50 & 0.553 &            &            & \checkmark & \checkmark & \checkmark &  \\
      0.00 & 0.75 & 0.576 &            &            & \checkmark & \checkmark & \checkmark &  \\
      0.00 & 1.00 & 0.601 &            &            & \checkmark & \checkmark &            &  \\
      \hline
      0.25 & 0.00 & 0.622 &            &            & \checkmark & \checkmark & \checkmark &  \\
      0.25 & 0.25 & 0.654 &            &            & \checkmark & \checkmark & \checkmark &  \\
      0.25 & 0.50 & 0.689 &            &            & \checkmark & \checkmark & \checkmark &  \\
      0.25 & 0.75 & 0.725 &            &            & \checkmark & \checkmark & \checkmark &  \\
      0.25 & 1.00 & 0.764 &            &            & \checkmark & \checkmark &            &  \\
      \hline
      0.50 & 0.00 & 0.791 &            &            & \checkmark & \checkmark & \checkmark &  \\
      0.50 & 0.25 & 0.844 &            &            & \checkmark & \checkmark & \checkmark &  \\
      0.50 & 0.50 & 0.901 &            &            & \checkmark & \checkmark & \checkmark &  \\
      0.50 & 0.75 & 0.963 &            &            & \checkmark & \checkmark & \checkmark &  \\
      0.50 & 1.00 & 1.030 &            &            & \checkmark & \checkmark &            &  \\
      \hline
      0.75 & 0.00 & 1.055 & \checkmark & \checkmark & \checkmark & \checkmark & \checkmark & $ \star $ \\
      0.75 & 0.25 & 1.146 & \checkmark & \checkmark & \checkmark & \checkmark & \checkmark & $ \star $ \\
      0.75 & 0.50 & 1.247 & \checkmark &            & \checkmark & \checkmark & \checkmark &  \\
      0.75 & 0.75 & 1.356 &            &            & \checkmark & \checkmark & \checkmark &  \\
      0.75 & 1.00 & 1.472 &            &            & \checkmark & \checkmark & \checkmark &  \\
      \hline
      1.00 & 0.00 & 1.461 & \checkmark & \checkmark &            & \checkmark & \checkmark &  \\
      1.00 & 0.25 & 1.603 & \checkmark & \checkmark &            & \checkmark & \checkmark &  \\
      1.00 & 0.50 & 1.750 & \checkmark & \checkmark & \checkmark & \checkmark & \checkmark & $ \star $ \\
      1.00 & 0.75 & 1.895 & \checkmark & \checkmark & \checkmark & \checkmark & \checkmark & $ \star $ \\
      1.00 & 1.00 & 2.032 & \checkmark & \checkmark & \checkmark & \checkmark & \checkmark & $ \star $ \\
    \end{tabular}
  \end{ruledtabular}
\end{table}
\subsubsection{Systematic behavior}
\label{sec:gap_systematic}
\par
Next, we show the systematic behavior of the pairing gaps and the charge radii of $ \mathrm{Ca} $, $ \mathrm{Sn} $, and $ \mathrm{Pb} $ isotopes.
In the following, we select four pairs of $ \tilde{h}_0 $ and $ \tilde{h}_{\urm{D}} $ among the five interactions shown with $ \star $ in Table~\ref{tab:strength_FaNDF0}.
We also use the volume-type pairing ($ \tilde{h}_0 = 0.00 $ and $ \tilde{h}_{\urm{D}} = 0.00 $)
as the representative of the conventional delta pairing interaction
and the interaction with $ \tilde{h}_0 = 0.00 $ and $ \tilde{h}_{\urm{D}} = 0.75 $
to see the effect of the Fayans-specific term ($ \tilde{h}_{\urm{D}} $ term).
Note that we confirmed that all the pairing interactions tested hereinafter give $ \Delta_p^{\urm{calc}} = 0 $ for all the tested $ \mathrm{Ca} $, $ \mathrm{Sn} $, and $ \mathrm{Pb} $ isotopes.
\par
\begin{figure}[tb]
  \centering
  \includegraphics[width=1.0\linewidth]{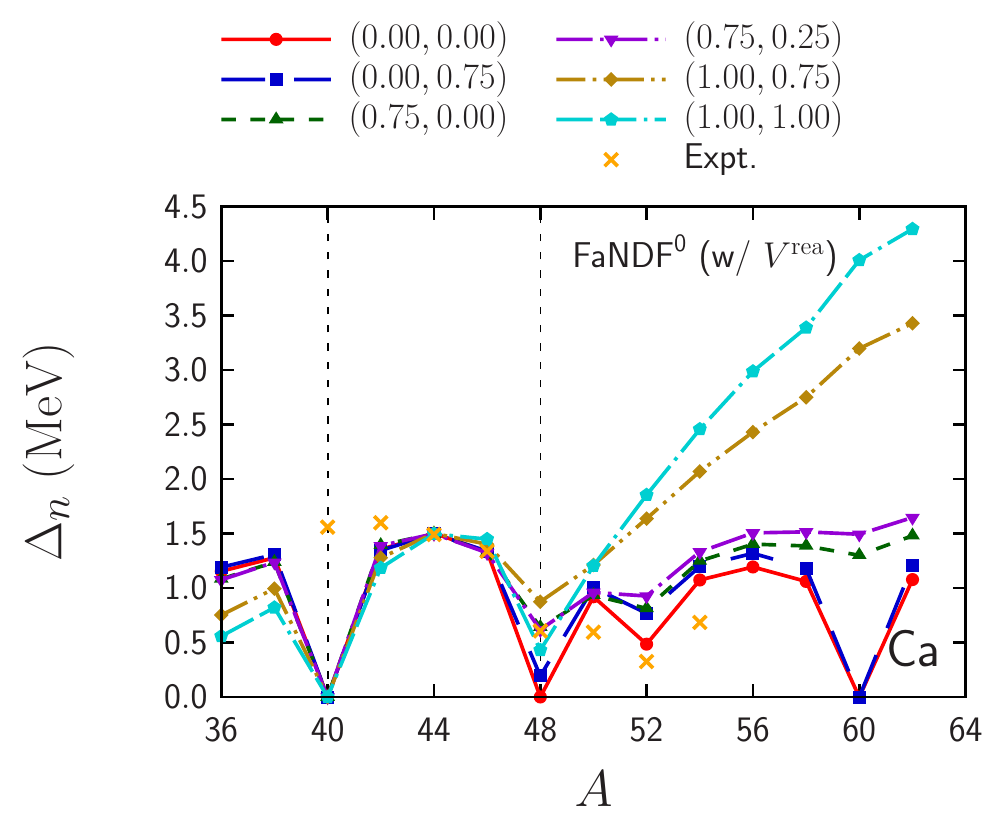}
  \caption{Neutron pairing gap $ \Delta_n $ for $ \mathrm{Ca} $ isotopes calculated with the selected $ \left( \tilde{h}_0, \tilde{h}_{\urm{D}} \right) $ pairs
    on top of the {\FaNDF} EDF.}
  \label{fig:FaNDF0_Deltan_020}
\end{figure}

Figure~\ref{fig:FaNDF0_Deltan_020} shows the neutron pairing gaps $ \Delta_n^{\urm{calc}} $ of $ \mathrm{Ca} $ isotopes.
For comparison, the experimental value $ \Delta_n^{\urm{ref}} $ obtained by Eq.~\eqref{eq:gap} is also shown.\footnote{
  It is well known that the simple three-point formula [Eq.~\eqref{eq:gap}] gives a nonzero value for a magic nucleus.}
It is confirmed that $ \Delta_n^{\urm{calc}} $ of $ \nuc{Ca}{40}{} $ vanishes in all the pairing interactions tested.
That of $ \nuc{Ca}{48}{} $ is small enough for the volume-type pairing,
but some of the pairing interactions give non-negligibly finite $ \Delta_n^{\urm{calc}} $,
which is in contrast to the known fact that $ \nuc{Ca}{48}{} $ is a doubly magic nucleus.
All the pairing interactions tested behave similarly in $ A \le 48 $.
In contrast, in the neutron-rich region ($ A > 48 $), different pairing interactions behave differently.
The volume-type pairing interaction shows magicity of $ \nuc{Ca}{52}{} $ and $ \nuc{Ca}{60}{} $,
while their magic character disappears with increasing $ \tilde{h}_0 $ and $ \tilde{h}_{\urm{D}} $,
and eventually, the pairing interaction of $ \tilde{h}_0 = 1.00 $ gives very large $ \Delta_n^{\urm{calc}} $ for $ A \ge 52 $.
\par
\begin{figure}[tb]
  \centering
  \includegraphics[width=1.0\linewidth]{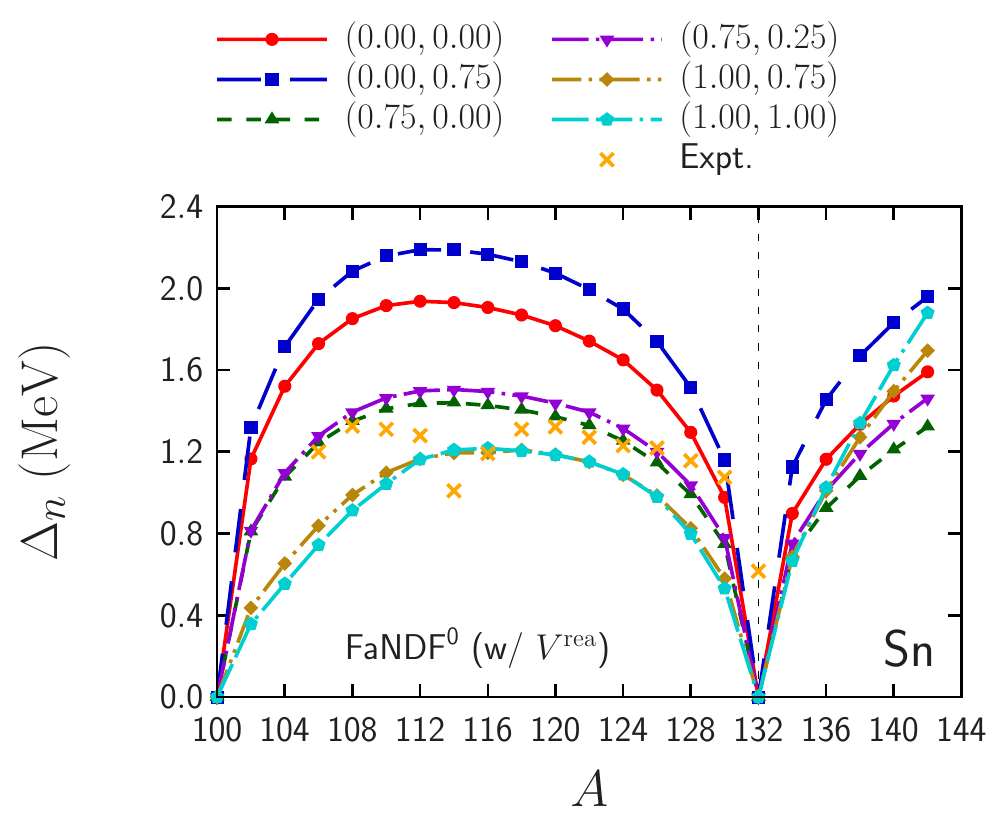}
  \caption{Same as Fig.~\ref{fig:FaNDF0_Deltan_020} but for $ \mathrm{Sn} $ isotopes.}
  \label{fig:FaNDF0_Deltan_050}
\end{figure}
\begin{figure}[tb]
  \centering
  \includegraphics[width=1.0\linewidth]{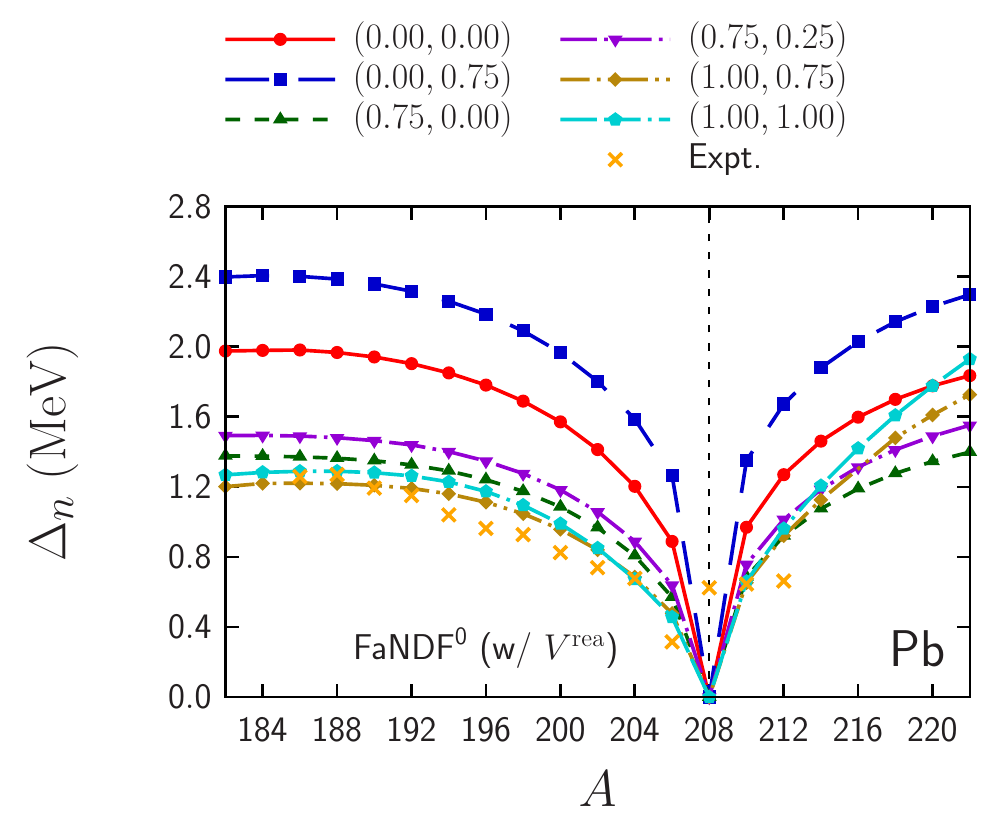}
  \caption{Same as Fig.~\ref{fig:FaNDF0_Deltan_020} but for $ \mathrm{Pb} $ isotopes.}
  \label{fig:FaNDF0_Deltan_082}
\end{figure}

Figures~\ref{fig:FaNDF0_Deltan_050} and \ref{fig:FaNDF0_Deltan_082} show the $ \Delta_n^{\urm{calc}} $ of $ \mathrm{Sn} $ and $ \mathrm{Pb} $ isotopes, respectively.
The interactions $ \left( 0.00, 0.00 \right) $ and $ \left( 0.00, 0.75 \right) $,
which did not comply with the selection conditions in Table~\ref{tab:strength_FaNDF0},
produce values of $ \Delta_n $ much larger than the other interactions.
Apart from the absolute values, all the pairing interactions give a similar $ A $ dependence,
although some small differences exist.
In $ \mathrm{Sn} $ isotopes,
all the interactions give almost the same behavior of the pairing gap between $ A = 114 $ and $ 132 $ if the pairing gap is normalized,
while the interactions with $ \tilde{h}_0 = 1.00 $ show differences from the others for $ A < 114 $ and $ A > 132 $.
Pairing interactions displayed in the figure cannot reproduce the dip shown by the experimental value $ \Delta_n^{\urm{ref}} $ in $ \nuc{Sn}{114}{} $.
In $ \mathrm{Pb} $ isotopes, the dependence of the gaps on $ A $ has a similar shape for all the interactions for $ A < 208 $.
For $ A > 208 $, the interactions with $ \tilde{h}_0 = 1.00 $ show differences from the others.
\subsection{Charge radii}
\label{sec:charge_radii}
\par
In this section, we discuss the charge radii of $ \nuc{Ca}{44}{} $, $ \nuc{Sn}{120}{} $, and $ \nuc{Pb}{204}{} $.
The experimental charge radii of these nuclei,
as well as of the reference nuclei $ \nuc{Ca}{40}{} $, $ \nuc{Sn}{132}{} $, and $ \nuc{Pb}{208}{} $,
are~\cite{
  Angeli2013At.DataNucl.DataTables99_69}
\begin{subequations}
  \begin{align}
    R_{\urm{ch}} \left( \nuc{Ca}{40}{} \right)
    & =
      3.4776 \, \mathrm{fm}, \\
    R_{\urm{ch}} \left( \nuc{Ca}{44}{} \right)
    & =
      3.5179 \, \mathrm{fm}, \\
    R_{\urm{ch}} \left( \nuc{Sn}{120}{} \right)
    & =
      4.6519 \, \mathrm{fm}, \\
    R_{\urm{ch}} \left( \nuc{Sn}{132}{} \right)
    & =
      4.7093 \, \mathrm{fm}, \\
    R_{\urm{ch}} \left( \nuc{Pb}{204}{} \right)
    & =
      5.4803 \, \mathrm{fm}, \\
    R_{\urm{ch}} \left( \nuc{Pb}{208}{} \right)
    & =
      5.5012 \, \mathrm{fm}.
  \end{align}
\end{subequations}
Accordingly, 
the relative differences of mean-square charge radius with respect to the reference magic nucleus are 
\begin{subequations}
  \begin{align}
    \delta_{\urm{expt}}^2 \left( \nuc{Ca}{44}{} \right)
    & =
      + 0.2819 \, \mathrm{fm}^2, \\    
    \delta_{\urm{expt}}^2 \left( \nuc{Sn}{120}{} \right)
    & =
      - 0.5373 \, \mathrm{fm}^2, \\    
    \delta_{\urm{expt}}^2 \left( \nuc{Pb}{204}{} \right)
    & =
      - 0.2295 \, \mathrm{fm}^2.
  \end{align}
\end{subequations}
\par
\begin{figure}[tb]
  \centering
  \includegraphics[width=1.0\linewidth]{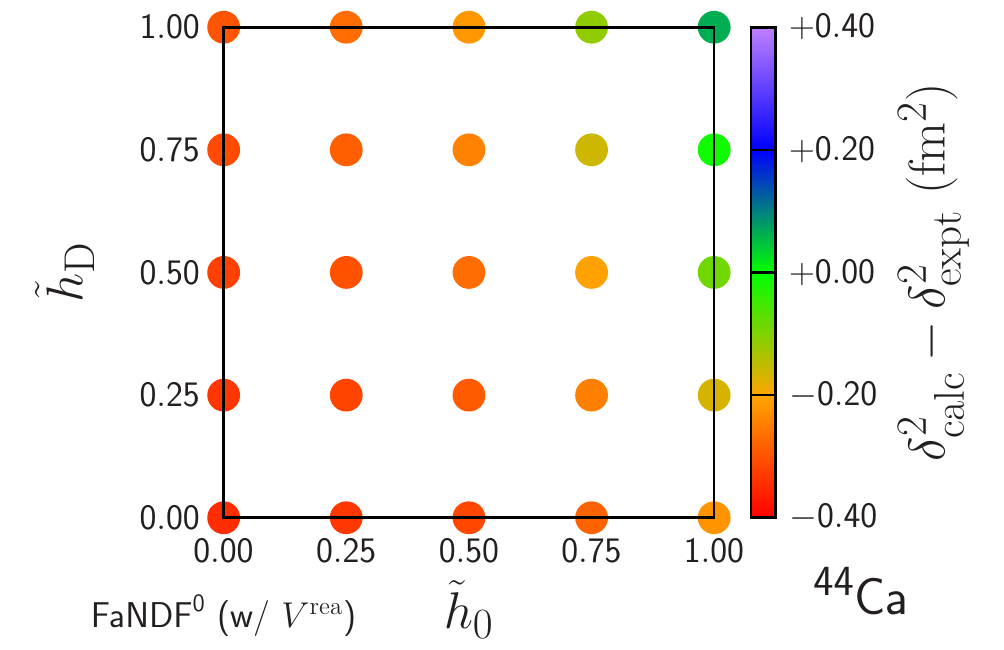}
  \caption{ 
    A comparison between the calculated and experimental relative differences $ \delta^2 $ of mean-square charge radii $ R_{\urm{ch}}^2 $ between $ \nuc{Ca}{44}{} $ 
    and the magic nucleus $ \nuc{Ca}{40}{} $. 
    The calculations are performed with various pairs of $ \tilde{h}_0 $ and $ \tilde{h}_{\urm{D}} $
    on top of the {\FaNDF} EDF.}
  \label{fig:FaNDF0_2D_Rch_020_044}
\end{figure}
\begin{figure}[tb]
  \centering
  \includegraphics[width=1.0\linewidth]{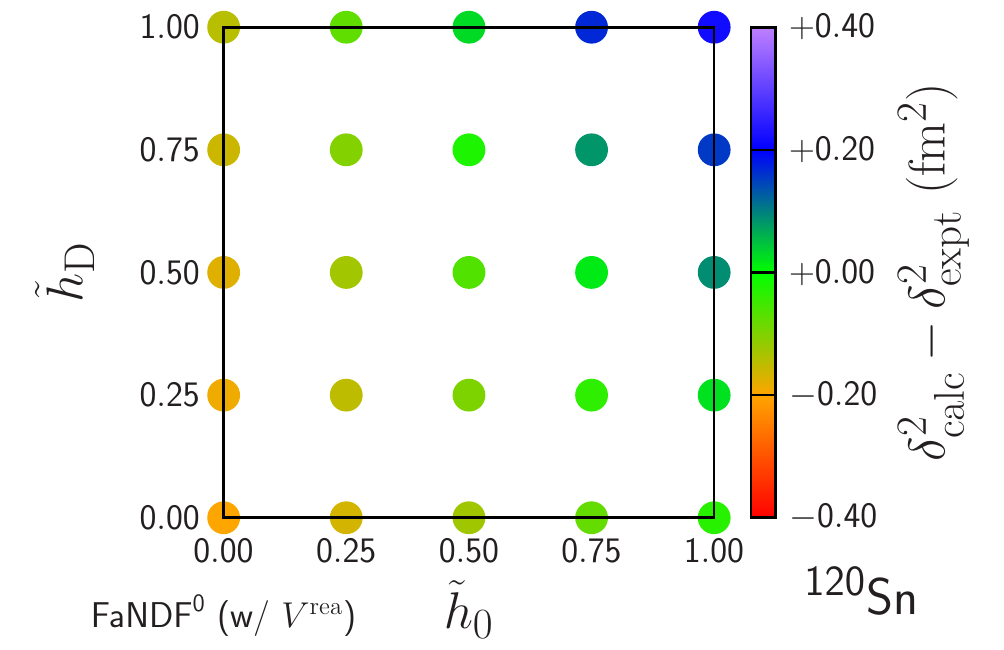}
  \caption{
    Same as Fig.~\ref{fig:FaNDF0_2D_Rch_020_044},
    but for $ R_{\urm{ch}}^2 $ of $ \nuc{Sn}{120}{} $ 
    with respect to the magic nucleus $ \nuc{Sn}{132}{} $.}
  \label{fig:FaNDF0_2D_Rch_050_120}
\end{figure}
\begin{figure}[tb]
  \centering
  \includegraphics[width=1.0\linewidth]{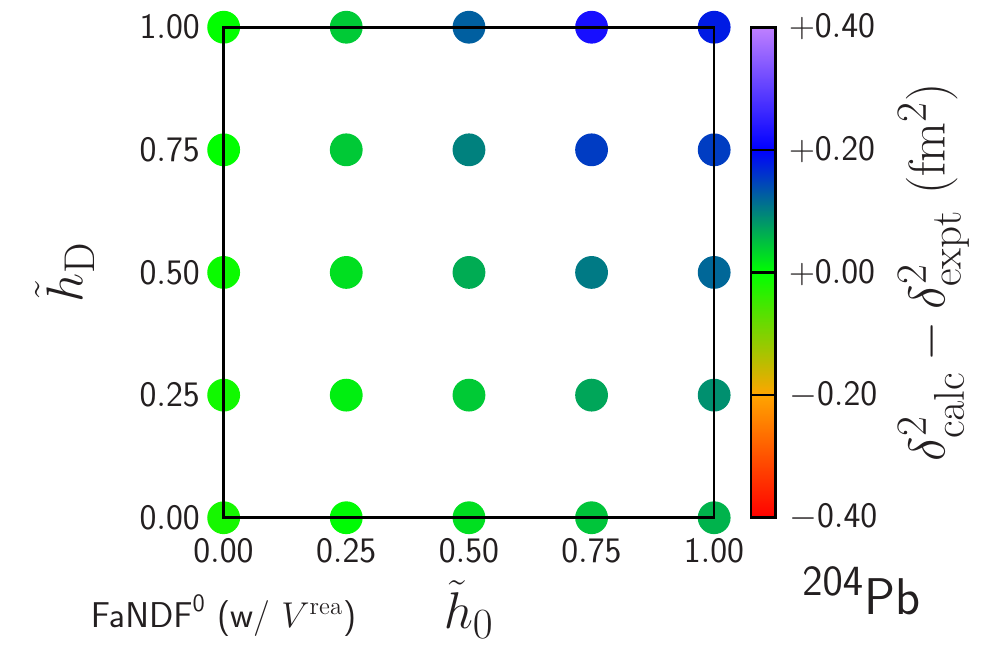}
  \caption{
    Same as Fig.~\ref{fig:FaNDF0_2D_Rch_020_044},
    but for $ R_{\urm{ch}}^2 $ of $ \nuc{Pb}{204}{} $ 
    with respect to the magic nucleus $ \nuc{Pb}{208}{} $.}
  \label{fig:FaNDF0_2D_Rch_082_204}
\end{figure}

Figures~\ref{fig:FaNDF0_2D_Rch_020_044}, \ref{fig:FaNDF0_2D_Rch_050_120}, and \ref{fig:FaNDF0_2D_Rch_082_204} show the differences
$ \delta_{\urm{calc}}^2 \left( \nuc{Ca}{44}{} \right) - \delta_{\urm{expt}}^2 \left( \nuc{Ca}{44}{} \right) $,
$ \delta_{\urm{calc}}^2 \left( \nuc{Sn}{120}{} \right) - \delta_{\urm{expt}}^2 \left( \nuc{Sn}{120}{} \right) $ and 
$ \delta_{\urm{calc}}^2 \left( \nuc{Pb}{204}{} \right) - \delta_{\urm{expt}}^2 \left( \nuc{Pb}{204}{} \right) $ respectively,
as functions of $ \tilde{h}_0 $ and $ \tilde{h}_{\urm{D}} $.
Different pairing interactions give different values of $ \delta^2 $.
One can notice that, in all three figures, 
$ \delta_{\urm{calc}}^2 - \delta_{\urm{expt}}^2 $
tends to increase as a function of $ \tilde{h}_0 $ for a fixed value of $ \tilde{h}_{\urm{D}} $,
and to increase (more mildly) as a function of $ \tilde{h}_{\urm{D}} $ for a fixed value of $ \tilde{h}_0 $. 
However, the values obtained for $ \left( \tilde{h}_0, \tilde{h}_{\urm{D}} \right) = \left( 0.00, 0.00 \right) $
are quite different in the three cases:
The value of $ \delta^2 $ is strongly underestimated in $ \nuc{Ca}{44}{} $,
is slightly underestimated in $ \nuc{Sn}{120}{} $, and is reproduced in $ \nuc{Pb}{204}{} $.
As a consequence, the values of the pairs that yield the best agreement are different in the three cases.
One notices that the case of $ \nuc{Ca}{44}{} $ is particularly selective,
because a good agreement (green points) is obtained only for
$ \tilde{h}_0 = 1.00 $ and $ \tilde{h}_{\urm{D}} \geq 0.5 $.
In the case of $ \nuc{Sn}{120}{} $ the best pairs tend to obey the condition
$ \tilde{h}_0  + \tilde{h}_{\urm{D}} \approx 1.00 $--$ 1.25 $,
while in $ \nuc{Pb}{204}{} $, one finds  $ \tilde{h}_0 \lesssim 0.25 $--$ 0.50 $.
\par
\begin{figure}[tb]
  \centering
  \includegraphics[width=1.0\linewidth]{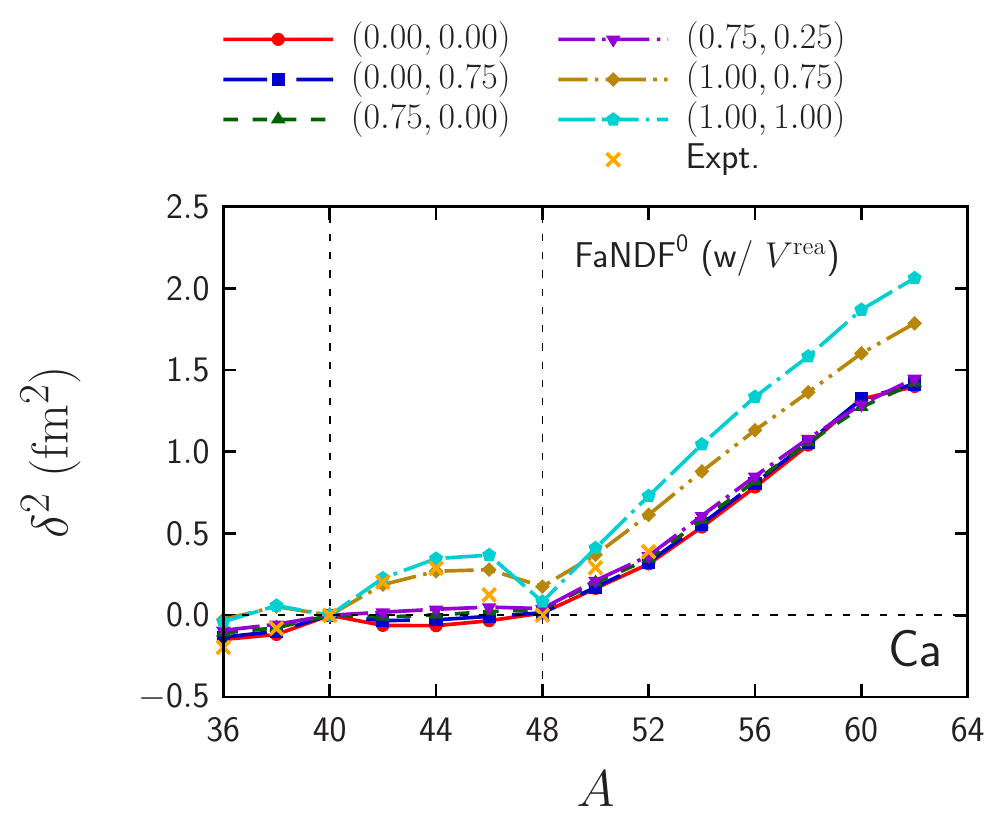}
  \caption{
    The difference $ \delta^2 $ between the mean-square charge radius calculated in
    $ \nuc{Ca}{A}{} $ isotopes and in $ \nuc{Ca}{40}{} $ as functions of $ A $.
    For comparison, the experimental data~\cite{
      Angeli2013At.DataNucl.DataTables99_69,
      GarciaRuiz2016Nat.Phys.12_594,
      Miller2019Nat.Phys.15_432}
    are shown.}
  \label{fig:FaNDF0_Rch_020}
\end{figure}
\begin{figure}[tb]
  \centering
  \includegraphics[width=1.0\linewidth]{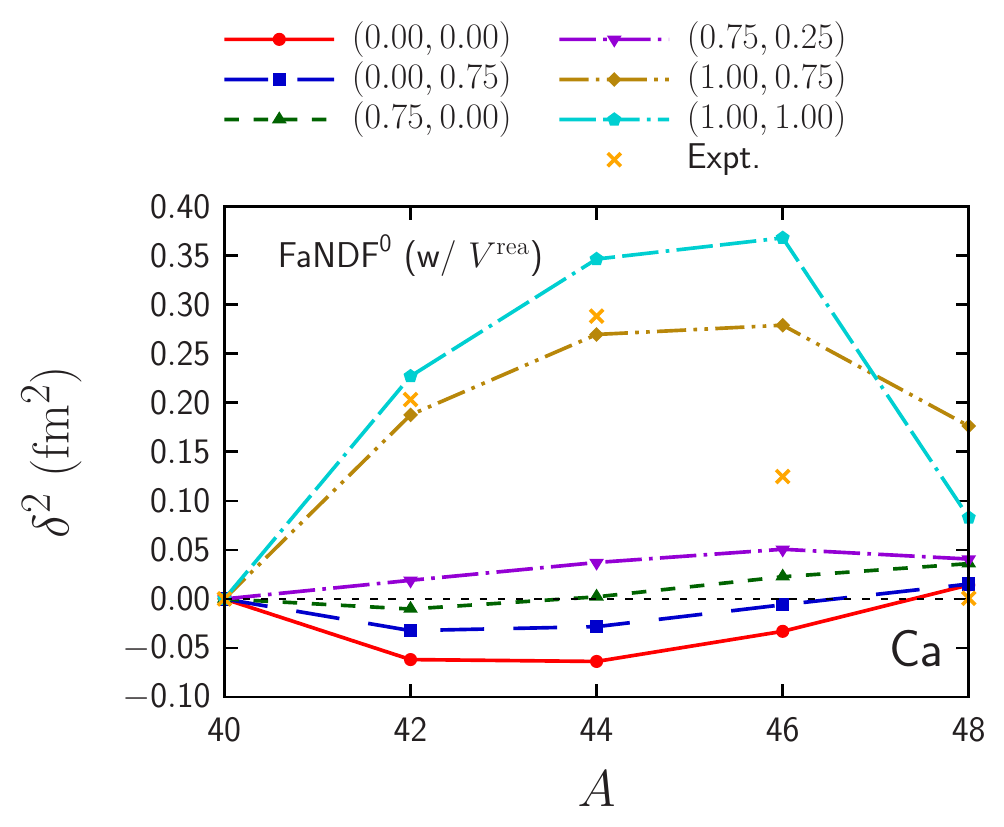}
  \caption{Same as Fig.~\ref{fig:FaNDF0_Rch_020} but only for $ 40 \le A \le 48 $.}
  \label{fig:FaNDF0_Rch_020_large}
\end{figure}

Figure~\ref{fig:FaNDF0_Rch_020} shows
$ \delta^2 \left( \nuc{Ca}{A}{} \right) $ of selected pairing interactions as functions of $ A $. 
To see the bell shape in detail,
the scale of the $ y $ axis, $ \delta^2 $, is enlarged for $ 40 \le A \le 48 $ in Fig.~\ref{fig:FaNDF0_Rch_020_large}.
The pairing interaction with $ \left( \tilde{h}_0, \tilde{h}_{\urm{D}} \right) = \left( 1.00, 0.75 \right) $ 
reproduces quite well the bell shape of $ \mathrm{Ca} $ isotopes,
in keeping with the above discussion for the case of $ \nuc{Ca}{44}{} $
and in agreement with several previous works~\cite{
  Reinhard2017Phys.Rev.C95_064328,
  Inakura2024Phys.Rev.C110_054315}.
The $ \left( 1.00, 1.00 \right) $ interaction even overestimates $ \delta^2 $ of $ \nuc{Ca}{42}{} $ and $ \nuc{Ca}{44}{} $.
The values of $ \delta^2 $ calculated with these two interactions increase quickly for $ A > 48 $
and overestimate the experimental value for $ \nuc{Ca}{52}{} $,
as well as for $ \nuc{Ca}{36}{} $ and $ \nuc{Ca}{38}{} $, as pointed out in Ref.~\cite{
  Inakura2024Phys.Rev.C110_054315}. 
The other interactions are not able to reproduce the bell shape, yielding $ \delta^2 \approx 0 $ for $ 40 \le A \le 48 $.
On the other hand, the increase of $ \delta^2 $ is milder for $ A > 48 $,
and the experimental value for $ \nuc{Ca}{52}{} $ is well reproduced. 
\par
\begin{figure}[tb]
  \centering
  \includegraphics[width=1.0\linewidth]{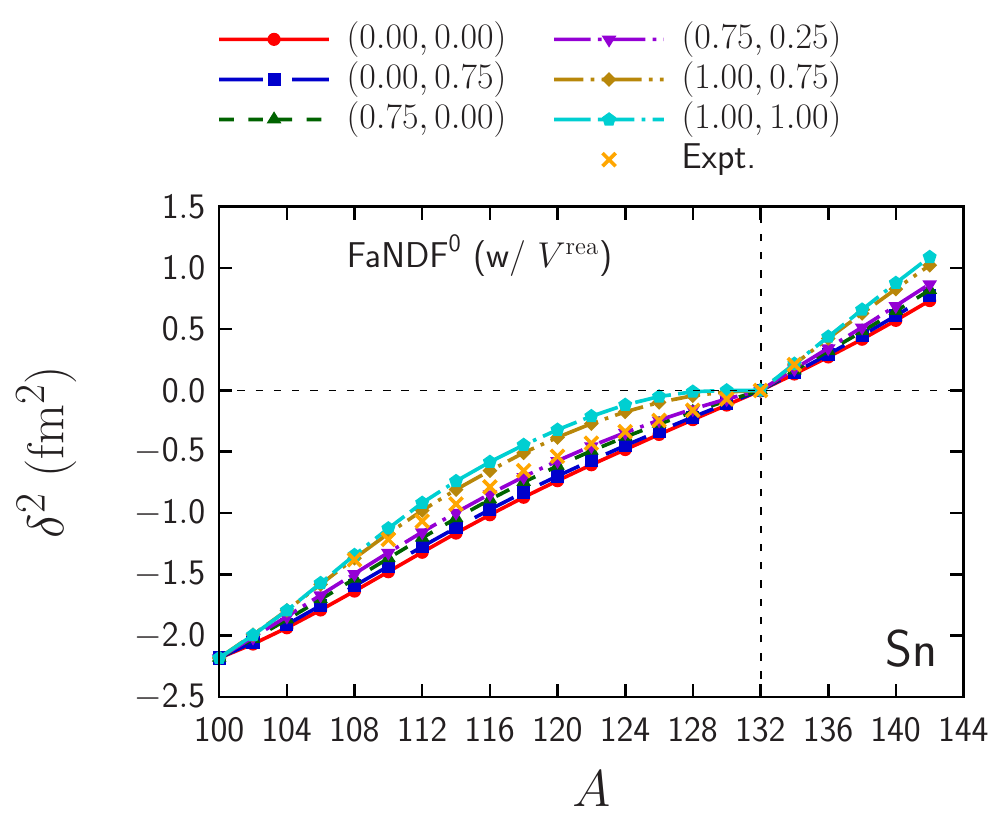}
  \caption{Same as Fig.~\ref{fig:FaNDF0_Rch_020} but for $ \mathrm{Sn} $ isotopes.
    For comparison, the experimental data~\cite{
      Angeli2013At.DataNucl.DataTables99_69,
      Gorges2019Phys.Rev.Lett.122_192502}
    are shown.}
  \label{fig:FaNDF0_Rch_050}
\end{figure}
\begin{figure}[tb]
  \centering
  \includegraphics[width=1.0\linewidth]{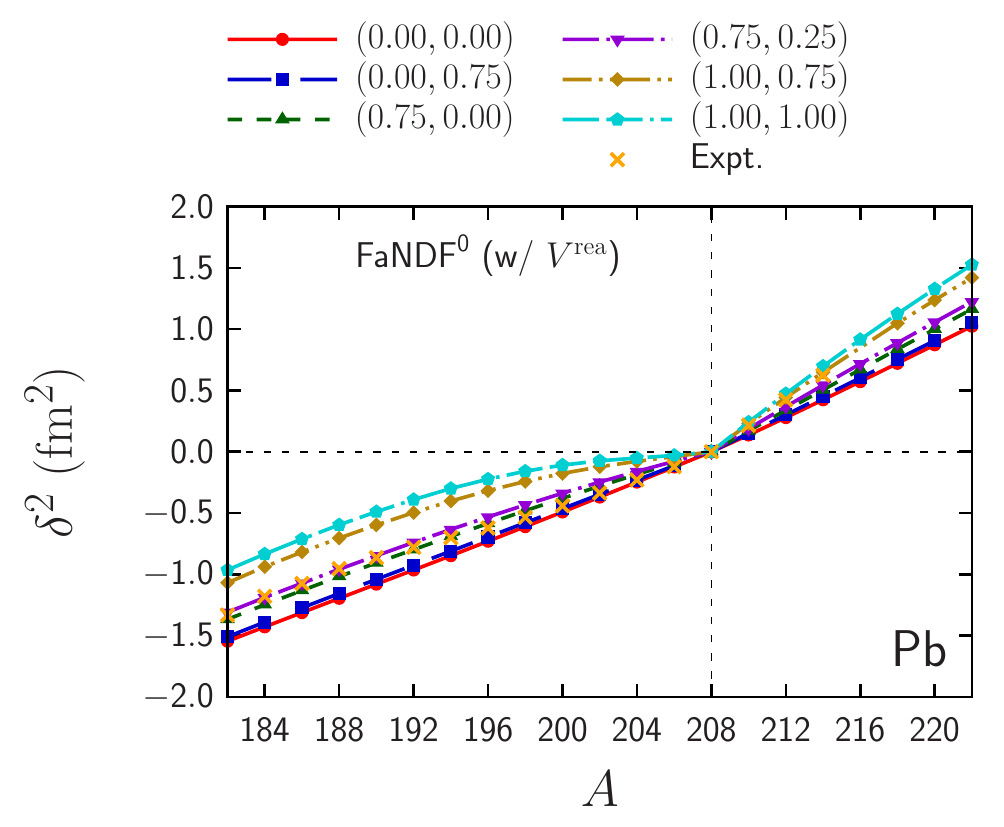}
  \caption{Same as Fig.~\ref{fig:FaNDF0_Rch_020} but for $ \mathrm{Pb} $ isotopes.
    For comparison, the experimental data~\cite{
      Angeli2013At.DataNucl.DataTables99_69}
    are shown.}
  \label{fig:FaNDF0_Rch_082}
\end{figure}

The trends observed in the $ \mathrm{Ca} $ chain are also visible in the case of $ \mathrm{Sn} $ and $ \mathrm{Pb} $,
although less pronounced
(see Figs.~\ref{fig:FaNDF0_Rch_050} and \ref{fig:FaNDF0_Rch_082}).
The two interactions with the largest values of $ \left( \tilde{h}_0, \tilde{h}_{\urm{D}} \right) $ tend to produce a more arched dependence of $ \delta^2 $ on $ A $ for 
$ A \le 132 $ and $ A \le 208 $ and a stronger slope for $ A > 132 $ and $ A > 208 $.
The experimental data are rather well reproduced overall by the
$ \left( \tilde{h}_0, \tilde{h}_{\urm{D}} \right) = \left( 0.75, 0.25 \right) $ interaction in $ \mathrm{Sn} $.
The same interaction works well for the $ \mathrm{Pb} $ chain,
although 
$ \left( \tilde{h}_0, \tilde{h}_{\urm{D}} \right) = \left( 0.00, 0.00 \right) $
and $ \left( \tilde{h}_0, \tilde{h}_{\urm{D}} \right) = \left( 0.00, 0.75 \right) $ 
give the best agreement with data for $ 196 < A < 208 $.
\clearpage
\subsection{Effect of the rearrangement potential}
\label{sec:rea}
\par
The present form of the Fayans pairing EDF was proposed in Ref.~\cite{
  Fayans1996Phys.Lett.B383_19},
where the repulsive term proportional to the square of the gradient of the total nuclear density
was introduced.
There, the radial form of the rearrangement potential was discussed in the case of $ \mathrm{Pb} $ isotopes,
finding that this potential gives a repulsive contribution to the mean field that pushes the proton density toward the exterior, increasing the charge radius.
Since then,  
the rearrangement potential $ V^{\urm{rea}} $ [Eq.~\eqref{eq:pot_rea}] for density-dependent pairing interactions has been seldom discussed in detail,
although it is known that its effects 
are essential, for instance, to produce the bell shape of $ \delta^2 $ in $ \mathrm{Ca} $ isotopes, as was pointed out in Ref.~\cite{
  Inakura2024Phys.Rev.C110_054315}.
In the following, we discuss the effects of the rearrangement potential, 
focusing on $ \nuc{Ca}{44}{} $ but considering also $ \mathrm{Sn} $ isotopes.
We study how the radial behavior of $ V^{\urm{rea}} $ depends on the  pairing interaction,
and show that the effects of $ V^{\urm{rea}} $ cannot be mimicked by simply adjusting the overall pairing strength.
\subsubsection{Systematic calculations}
\begin{figure}[h]
  \centering
  \includegraphics[width=1.0\linewidth]{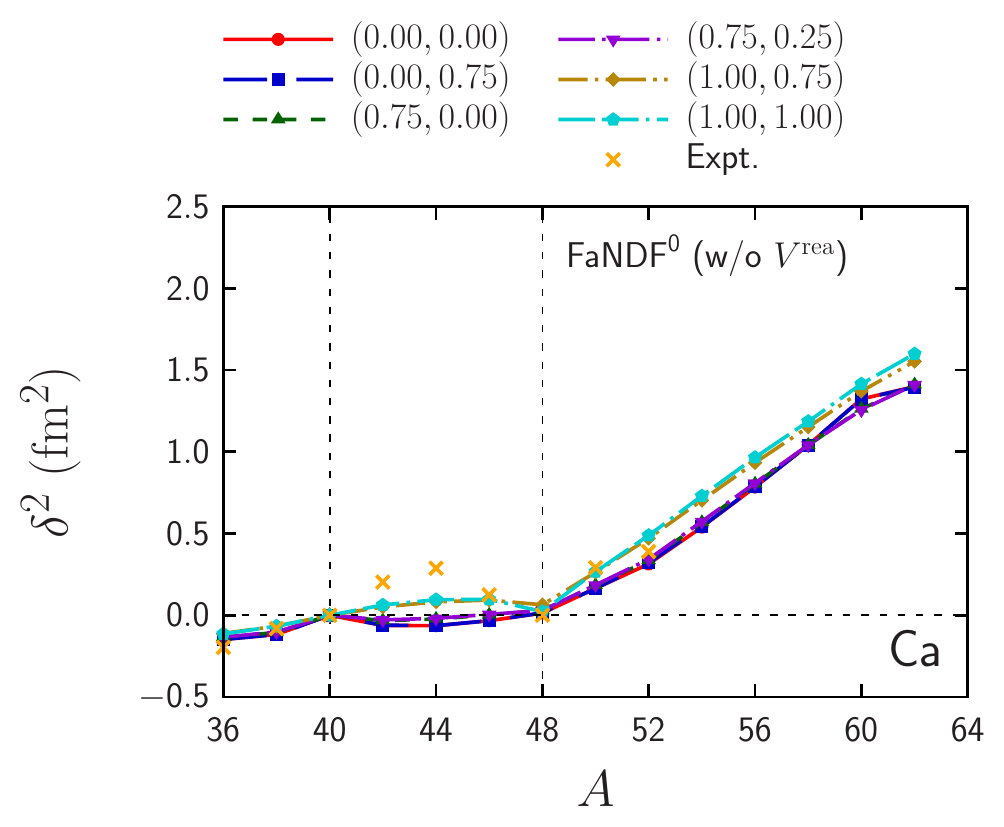}
  \caption{Same as Fig.~\ref{fig:FaNDF0_Rch_020} but without considering the rearrangement potential $ V^{\urm{rea}} $.}
  \label{fig:FaNDF0_woVrea_Rch_020}
\end{figure}
\begin{figure}[h]
  \centering
  \includegraphics[width=1.0\linewidth]{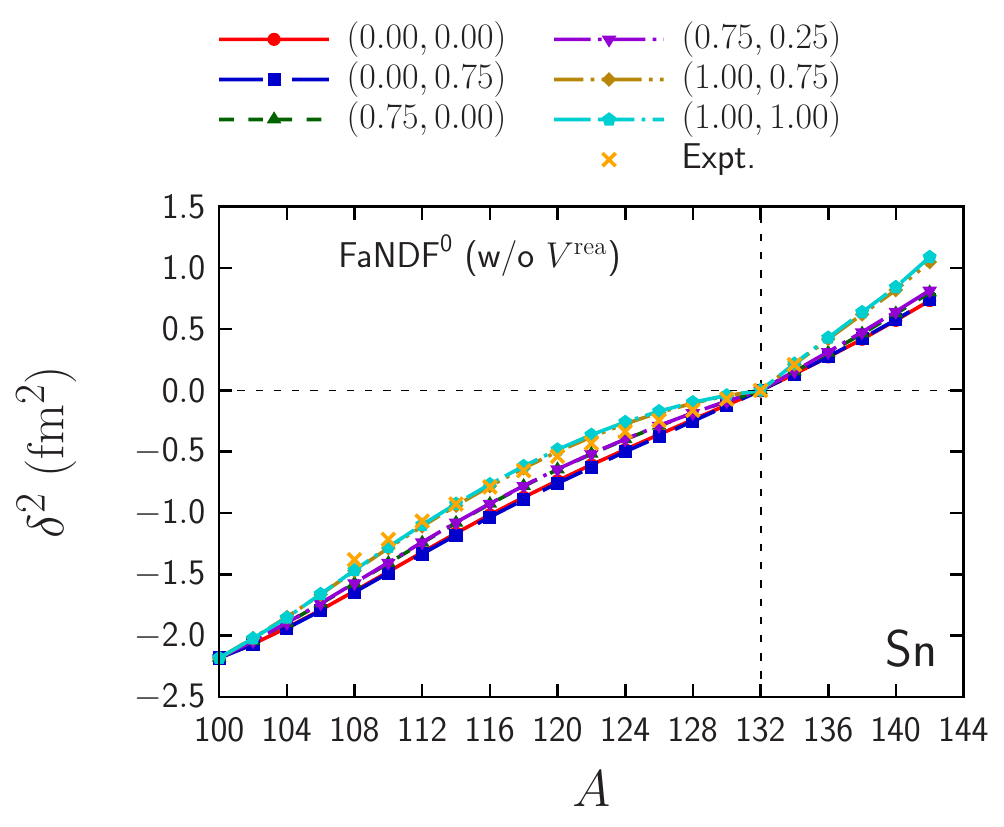}
  \caption{Same as Fig.~\ref{fig:FaNDF0_Rch_050} but without considering the rearrangement potential $ V^{\urm{rea}} $.}
  \label{fig:FaNDF0_woVrea_Rch_050}
\end{figure}
\begin{figure}[h]
  \centering
  \includegraphics[width=1.0\linewidth]{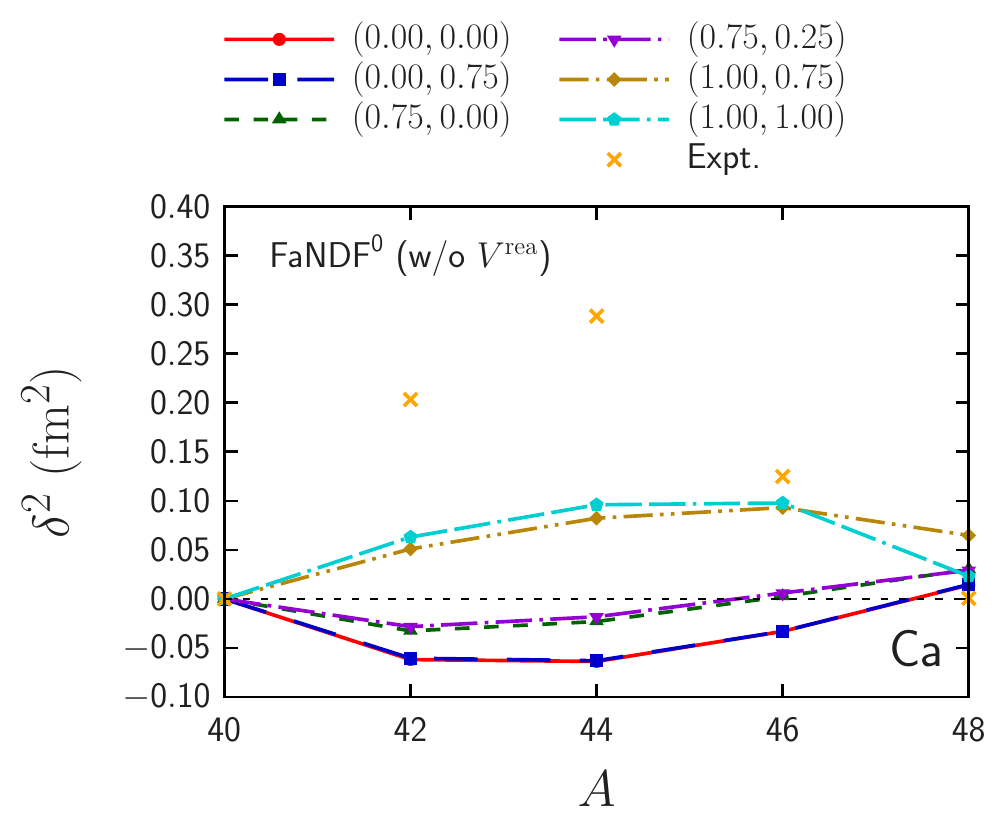}
  \caption{Same as Fig.~\ref{fig:FaNDF0_woVrea_Rch_020} but only for $ 40 \le A \le 48 $.}
  \label{fig:FaNDF0_woVrea_large_020}
\end{figure}
\begin{figure}[h]
  \centering
  \includegraphics[width=1.0\linewidth]{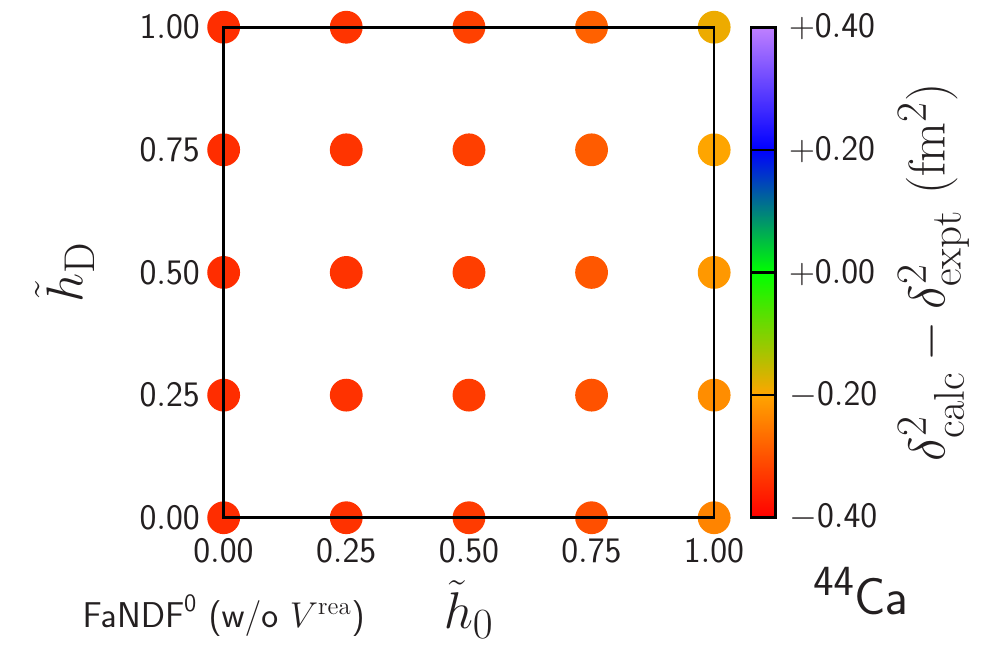}
  \caption{Same as Fig.~\ref{fig:FaNDF0_2D_Rch_020_044} but without considering the rearrangement potential $ V^{\urm{rea}} $.}
  \label{fig:FaNDF0_woVrea_2D_Rch_020_044}
\end{figure}
\begin{figure}[h]
  \centering
  \includegraphics[width=1.0\linewidth]{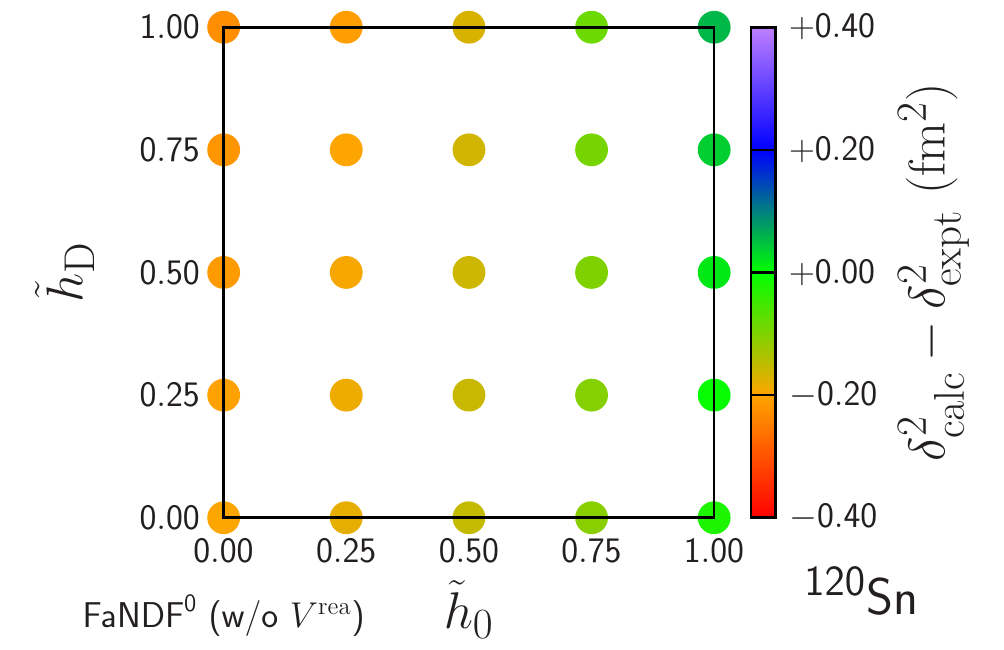}
  \caption{Same as Fig.~\ref{fig:FaNDF0_2D_Rch_050_120} but without considering the rearrangement potential $ V^{\urm{rea}} $.}
  \label{fig:FaNDF0_woVrea_2D_Rch_050_120}
\end{figure}
\begin{figure}[tb]
  \centering
  \includegraphics[width=1.0\linewidth]{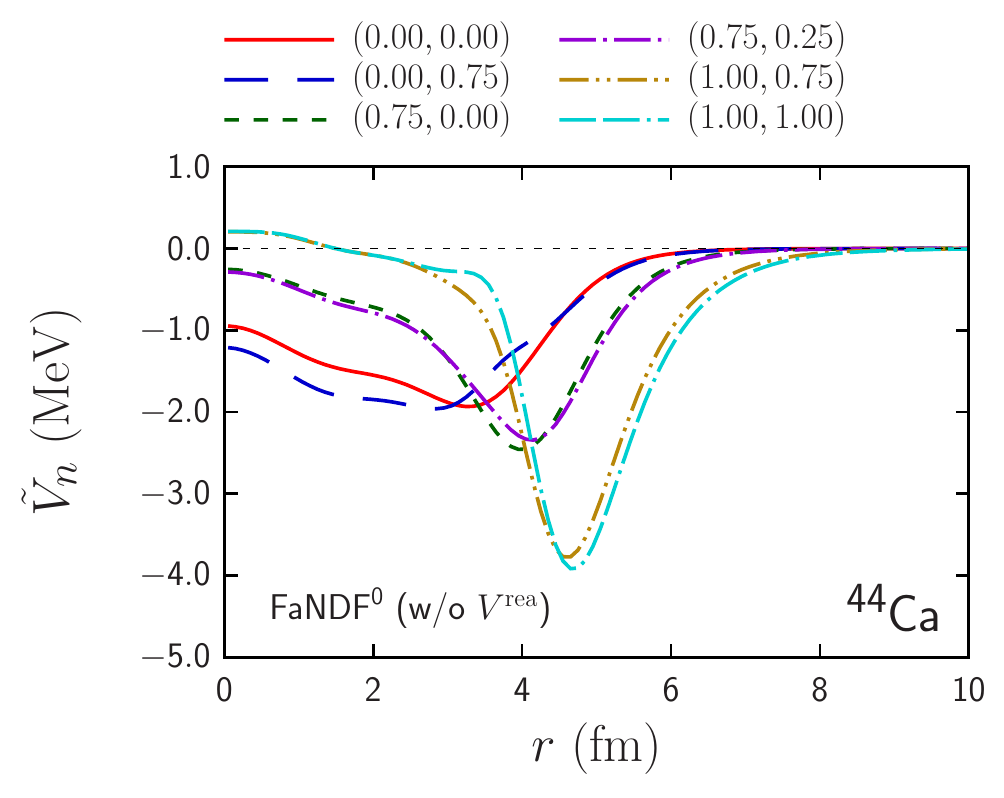}
  \caption{Neutron pairing potential $ \tilde{V}_n $ of $ \nuc{Ca}{44}{} $
    obtained by the selected $ \left( \tilde{h}_0, \tilde{h}_{\urm{D}} \right) $ pairs
    on top of the {\FaNDF} EDF.
    The rearrangement potential is not considered in this calculation.}
  \label{fig:FaNDF0_woVrea_Vpair_020_044}
\end{figure}
\begin{figure}[tb]
  \centering
  \includegraphics[width=1.0\linewidth]{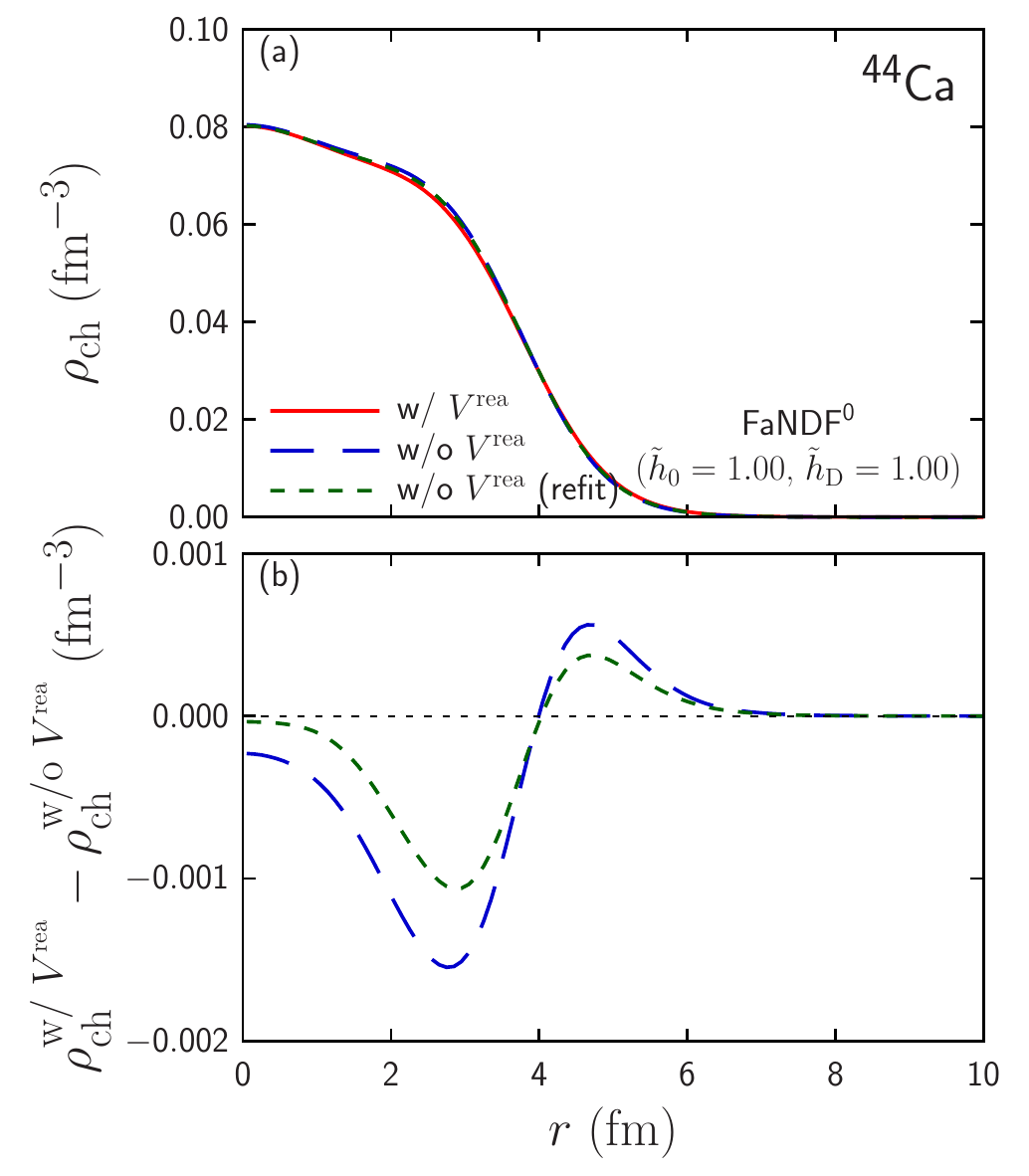}
  \caption{(a) Charge density $ \rho_{\urm{ch}} $ of $ \nuc{Ca}{44}{} $ calculated with $ \tilde{h}_0 = \tilde{h}_{\urm{D}} = 1.00 $ with (solid line) and without (long-dashed line) the rearrangement potential.
    The charge density obtained without the rearrangement potential with the refitted strength is also shown as a dashed line.
    (b) The difference between $ \rho_{\urm{ch}} $ obtained with and without the rearrangement potential.
    The same lines are used as in panel (a).}
  \label{fig:FaNDF0_charge-den_020_044_refit}
\end{figure}
\par
We perform a series of calculations without including $ V^{\urm{rea}} $.
The resulting values of $ \delta^2 $ for $ \mathrm{Ca} $ and $ \mathrm{Sn} $ isotopes
are, respectively, shown in Figs.~\ref{fig:FaNDF0_woVrea_Rch_020} and \ref{fig:FaNDF0_woVrea_Rch_050}.
The scale of the $ y $ axis for $ \delta^2 $ is enlarged for $ \mathrm{Ca} $ isotopes with $ 40 \le A \le 48 $ in Fig.~\ref{fig:FaNDF0_woVrea_large_020}.
Comparing these results with the corresponding calculations carried out including $ V^{\urm{rea}} $,
previously shown in Figs.~\ref{fig:FaNDF0_Rch_020}--\ref{fig:FaNDF0_Rch_050}, it is apparent that the bell-shaped
dependence of charge radii is much less pronounced in the absence of $ V^{\urm{rea}} $.
The agreement 
with the data in $ \nuc{Ca}{42}{} $--$ \nuc{Ca}{46}{} $ isotopes is then spoiled, while it is considerably improved in the case of $ \mathrm{Sn} $ isotopes.
Comparing Fig.~\ref{fig:FaNDF0_woVrea_2D_Rch_020_044} with 
Fig.~\ref{fig:FaNDF0_2D_Rch_020_044} and 
Fig.~\ref{fig:FaNDF0_woVrea_2D_Rch_050_120} with Fig.~\ref{fig:FaNDF0_2D_Rch_050_120},
one notices that, without $ V^{\urm{rea}} $,
$ \delta^2_{\urm{calc}} $ hardly depends on $ \tilde{h}_{\urm{D}} $.
These results originate from the fact that
in this case the only quantity which can make a difference among different choices of
$ \tilde{h}_0 $ and $ \tilde{h}_{\urm{D}} $ is the pair potential itself,
and $ \tilde{V}_n $ is rather insensitive to $ \tilde{h}_{\urm{D}} $, as can be seen in Fig.~\ref{fig:FaNDF0_woVrea_Vpair_020_044}.
\clearpage
\subsubsection{Pairing and rearrangement potentials of $ \nuc{Ca}{44}{} $}
\begin{figure}[tb]
  \centering
  \includegraphics[width=1.0\linewidth]{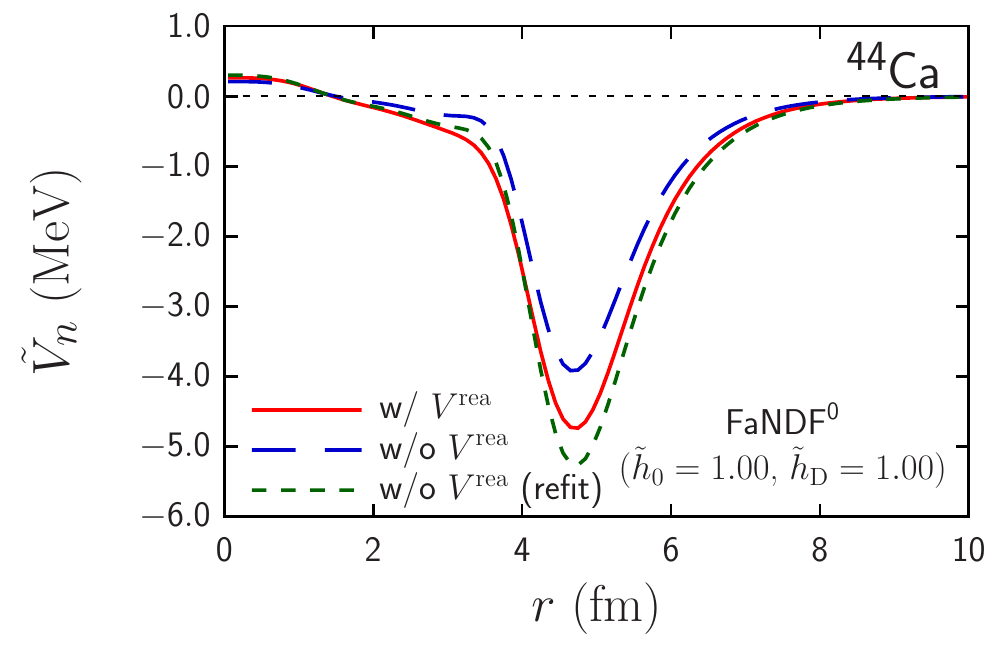}
  \caption{Neutron pairing potential $ \tilde{V}_n $ of $ \nuc{Ca}{44}{} $ calculated with $ \tilde{h}_0 = \tilde{h}_{\urm{D}} = 1.00 $ with and without the rearrangement potential shown by solid and long-dashed lines, respectively.
    The pairing potential obtained without the rearrangement potential with the refitted strength is also shown by the dashed line.}
  \label{fig:FaNDF0_Vpair_020_044_100_100_refit}
\end{figure}
\par
From now on, we focus on the case of $ \nuc{Ca}{44}{} $.
We mostly consider 
the case corresponding to the parameters $ \tilde{h}_0 = \tilde{h}_{\urm{D}} = 1.00 $, which enhance the effects of $ V^{\urm{rea}}$.
The most important effect of $ V^{\urm{rea}} $ is the modification of the charge density. 
As is shown in Fig.~\ref{fig:FaNDF0_charge-den_020_044_refit}(b),
some part of the charge density is moved from the interior 
to the exterior of the nucleus, increasing the charge radius.
This change must be related to a modification of the mean field and of the pairing field, 
with the  consequent change of the energy of the single-particle levels and of their radial dependence and occupation probabilities.
This is discussed in Sec.~\ref{sec:refit}.
\par
The pairing potentials obtained with and without $ V^{\urm{rea}} $ are compared in Fig.~\ref{fig:FaNDF0_Vpair_020_044_100_100_refit}.
Without $ V^{\urm{rea}} $, the magnitude of $ \tilde{V}_n $ is reduced by about $ 20 \, \% $,
while the radial dependence remains approximately the same.
\par
The rearrangement potential [Eq.~\eqref{eq:pot_rea}] can be divided into three terms:
\begin{subequations}
  \begin{align}
    V^{\urm{rea}}_1 \left( \ve{r} \right)
    & = 
      -
      \frac{\epsilon_{\urm{F}} f}{3 \rho_0^2}
      \tilde{h}_0 \gamma 
      \alpha^{\gamma - 1}
      \sum_{q'}
      h_{q'}
      \left[
      \tilde{\rho}_{q'} \left( \ve{r} \right)
      \right]^2, \\
    V^{\urm{rea}}_2 \left( \ve{r} \right)
    & = 
      \frac{2 \epsilon_{\urm{F}} f}{3 \rho_0^2}
      \tilde{h}_{\urm{D}}
      r_{\urm{s}}^2
      \laplace \alpha
      \sum_{q'}
      h_{q'}
      \left[
      \tilde{\rho}_{q'} \left( \ve{r} \right)
      \right]^2, \\
    V^{\urm{rea}}_3 \left( \ve{r} \right)
    & = 
      \frac{4 \epsilon_{\urm{F}} f}{3 \rho_0^2}
      \tilde{h}_{\urm{D}}
      r_{\urm{s}}^2
      \nabla \alpha
      \cdot
      \sum_{q'}
      h_{q'}
      \tilde{\rho}_{q'} \left( \ve{r} \right)
      \nabla
      \tilde{\rho}_{q'} \left( \ve{r} \right), \\
    V^{\urm{rea}} \left( \ve{r} \right)
    & =
      V^{\urm{rea}}_1 \left( \ve{r} \right)
      +
      V^{\urm{rea}}_2 \left( \ve{r} \right)
      +
      V^{\urm{rea}}_3 \left( \ve{r} \right).
  \end{align}
\end{subequations}
The rearrangement potential for protons is identical to that for neutrons 
even if the proton pairing field vanishes.
The first term $ V^{\urm{rea}}_1 $ depends only on $ \tilde{h}_0 $
and the others on $ \tilde{h}_{\urm{D}} $.
The first term is always nonnegative since $ f < 0 $ holds.
Therefore, the rearrangement potential is never attractive if $ \tilde{h}_{\urm{D}} = 0.00 $.
\begin{figure}[h]
  \centering
  \includegraphics[width=1.0\linewidth]{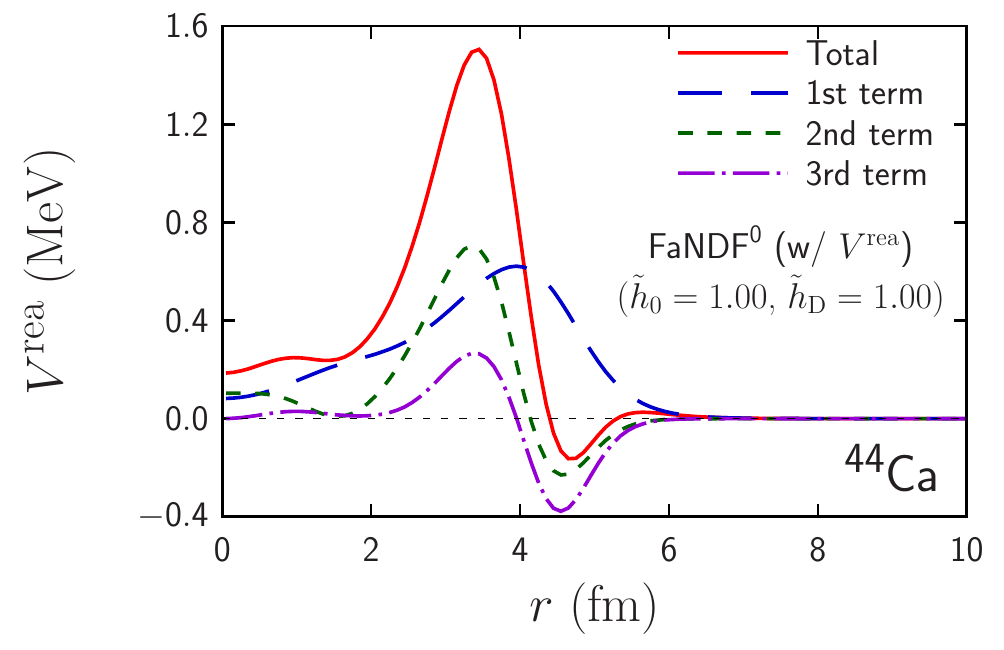}
  \caption{Rearrangement potential $ V^{\urm{rea}} $ of $ \nuc{Ca}{44}{} $
    obtained by $ \tilde{h}_0 = \tilde{h}_{\urm{D}} = 1.00 $.
    The breakdown of the rearrangement potential is also shown.}
  \label{fig:FaNDF0_Vrea_020_044_100_100}
\end{figure}
\begin{figure}[h]
  \centering
  \includegraphics[width=1.0\linewidth]{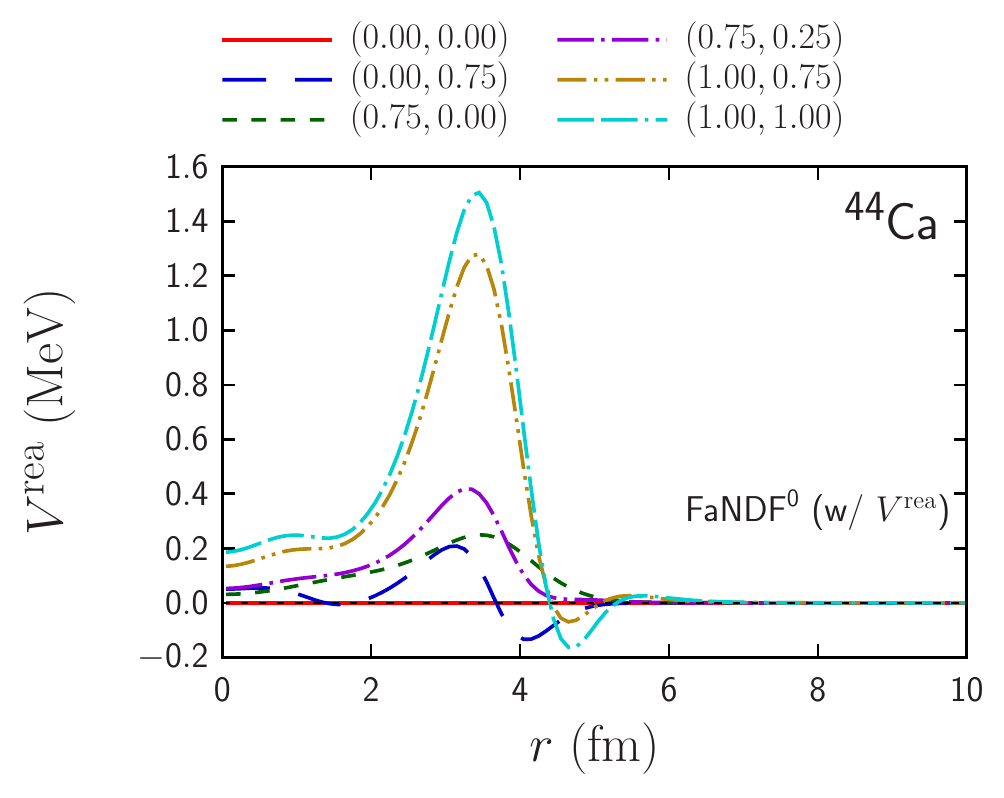}
  \caption{Rearrangement potential $ V^{\urm{rea}} $ of $ \nuc{Ca}{44}{} $
    obtained by the selected $ \left( \tilde{h}_0, \tilde{h}_{\urm{D}} \right) $ pairs
    on top of the {\FaNDF} EDF.}
  \label{fig:FaNDF0_Vrea_020_044}
\end{figure}
\begin{figure}[h]
  \centering
  \includegraphics[width=1.0\linewidth]{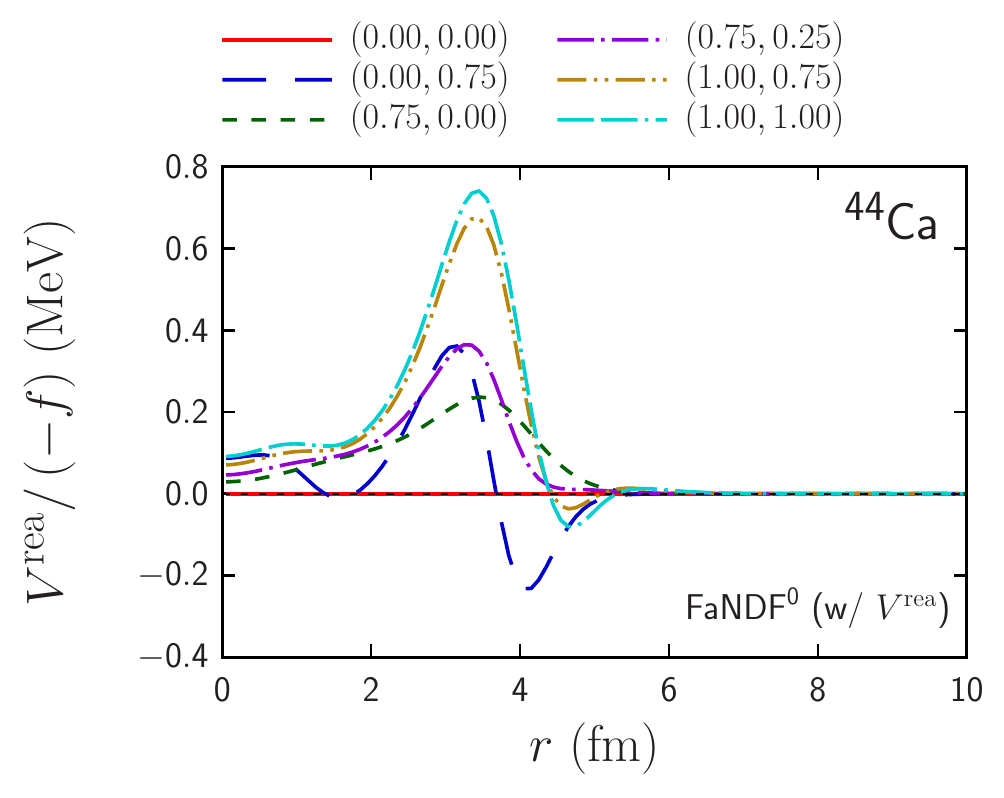}
  \caption{Rearrangement potential $ V^{\urm{rea}} $ of $ \nuc{Ca}{44}{} $
    normalized by $ -f $.}
  \label{fig:FaNDF0_Vrea_020_044_normalized}
\end{figure}
\par
Figure~\ref{fig:FaNDF0_Vrea_020_044_100_100} shows the different components  
in the case of $ \nuc{Ca}{44}{} $, using the interaction $ \tilde{h}_0 = \tilde{h}_{\urm{D}} = 1.00 $ that produces the largest rearrangement potential. 
The three terms are positive in the interior and produce a strong repulsive peak for $ r \approx 3.5 \, \mathrm{fm} $.
The first term remains positive on the surface, while the second and third terms become negative.
As a consequence, the total $ V^{\urm{rea}} $ becomes slightly attractive for $ r \approx 4.5 \, \mathrm{fm} $.
It is seen that both parameters $ \tilde{h}_0 $ and $ \tilde{h}_{\urm{D}} $
are important in determining the spatial dependence of the rearrangement term
and
give the repulsive nature of the rearrangement potential with almost the same amount.
The correspondence between the radial dependence of $ V^{\urm{rea}} $ and the modifications of the charge density shown in Fig.~\ref{fig:FaNDF0_charge-den_020_044_refit} is apparent.
\par
The rearrangement potential $ V^{\urm{rea}} $ of $ \nuc{Ca}{44}{} $ is plotted in Fig.~\ref{fig:FaNDF0_Vrea_020_044} for different values of $ \left( \tilde{h}_0, \tilde{h}_{\urm{D}} \right) $.
Even if $ \tilde{h}_0 = 0.00 $,
$ V^{\urm{rea}} $ keeps its repulsive character for $ \tilde{h}_{\urm{D}} = 0.75 $.
Different values of $ \tilde{h}_{\urm{D}} $ with the same $ \tilde{h}_0 $ yield a different shape of the rearrangement potential,
whereas in the $ \tilde{h}_0 = 1.00 $ case, two different $ \tilde{h}_{\urm{D}} $ give a similar shape.
\par
The magnitude of $ V^{\urm{rea}} $ is much enhanced for 
$ \left( \tilde{h}_0, \tilde{h}_{\urm{D}} \right) = \left( 1.00, 0.75 \right) $ and $ \left( 1.00, 1.00 \right) $.
This is partly due to the fact that
in this case the pairing strength $ f $ takes large values
for larger $ \tilde{h}_0 $,
as can be seen in Table~\ref{tab:strength_FaNDF0}.
However, even dividing the rearrangement potentials by $ -f > 0 $,
the interactions $ \left( \tilde{h}_0, \tilde{h}_{\urm{D}} \right) = \left( 1.00, 0.75 \right) $ and $ \left( 1.00, 1.00 \right) $
still produce the largest $ V^{\urm{rea}} $,
as shown in Fig.~\ref{fig:FaNDF0_Vrea_020_044_normalized}.
This is because, as discussed above,
$ \tilde{h}_0 $ and $ \tilde{h}_{\urm{D}} $ terms contribute almost equally to the repulsive nature of the rearrangement potential.
\subsubsection{Analysis based on refitted pairing strength}
\label{sec:refit}
\par
One may wonder if the role of the rearrangement potential is really unique, or if instead its effects can be mocked up by a different choice of the pairing strength.
In this section, we show that the latter possibility does not hold, and a large fraction of the enhancement of the charge radius emerges genuinely by the rearrangement potential only.
\par
Again, we take $ \tilde{h}_0 = \tilde{h}_{\urm{D}} = 1.00 $ as an example.
We have seen that the rearrangement potential pushes the density outwards in Fig.~\ref{fig:FaNDF0_charge-den_020_044_refit}(b) and,
in turn, enhances the pairing potential in Fig.~\ref{fig:FaNDF0_Vpair_020_044_100_100_refit},
increasing the neutron pairing gap in $ \nuc{Ca}{44}{} $ 
from $ 1.05 \, \mathrm{MeV} $ to $ 1.50 \, \mathrm{MeV} $ 
(see Table~\ref{tab:FaNDF0_020_044_proton}).
This change of the pairing gap also affects the values of the quasiparticle energy $ E_{\urm{qp}} $
and the occupation probability of the neutron $ 1f_{7/2} $ orbital
(estimated by the spatial integral of $ v^2 $),
which go from $ E_{\urm{qp}} = 1.558 \, \mathrm{MeV} $ and $ v^2 = 0.442 $ without $ V^{\urm{rea}} $ to $ 2.161 \, \mathrm{MeV} $ and $ 0.406 $ with $ V^{\urm{rea}} $,
as seen in Table~\ref{tab:FaNDF0_020_044_proton}.
\begin{figure}[h]
  \centering
  \includegraphics[width=1.0\linewidth]{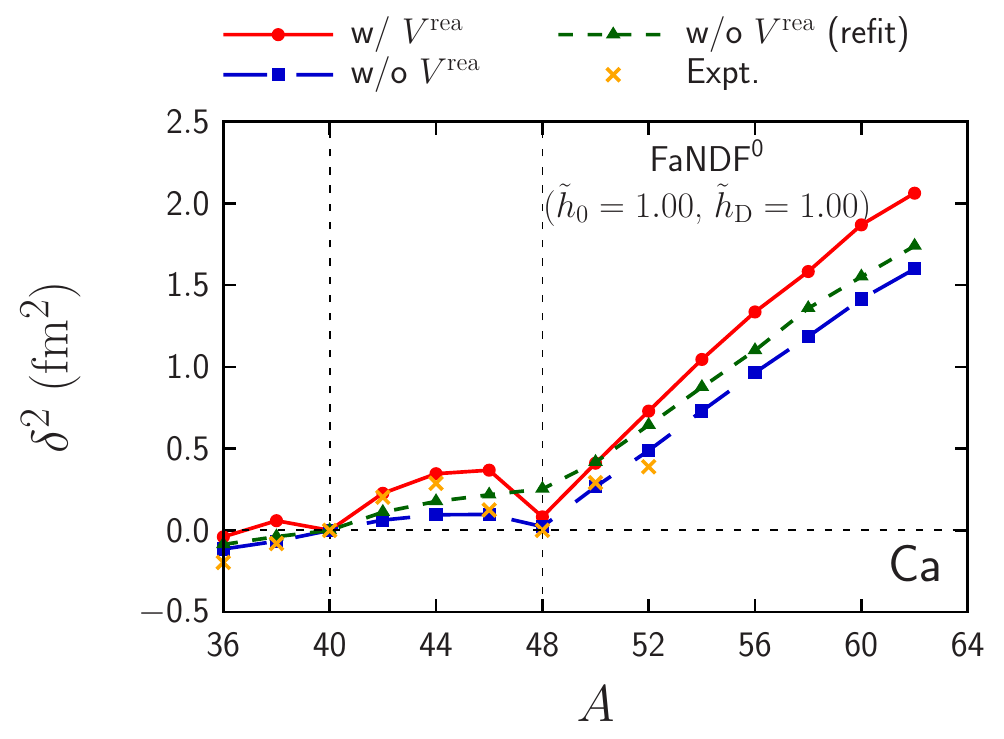}
  \caption{Same as Fig.~\ref{fig:FaNDF0_Rch_020}
    but only for $ \tilde{h}_0 = \tilde{h}_{\urm{D}} = 1.00 $.
    Calculation with the rearrangement potential (solid),
    calculation without the rearrangement potential (long-dashed),
    and 
    calculation without the rearrangement potential with the refitted strength (dashed) are shown.}
  \label{fig:FaNDF0_refit_Rch_020}
\end{figure}
\begingroup
\squeezetable
\begin{table}[tb]
  \centering
  \caption{Neutron $ 1f_{7/2} $ quasiparticle (QP) orbital energy,
    radius obtained by the $ v $ component,
    and its occupation ($ v^2 $),
    proton single-particle (SP) energy and its radius,
    and proton and neutron radii of $ \nuc{Ca}{44}{} $ obtained by
    $ \tilde{h}_0 = \tilde{h}_{\urm{D}} = 1.00 $.
    Energies and radii are, respectively, shown in $ \mathrm{MeV} $ and $ \mathrm{fm} $.}
  \label{tab:FaNDF0_020_044_proton}
  \begin{ruledtabular}
    \begin{tabular}{lllddd}
      \multicolumn{3}{c}{Orbital} & \multicolumn{1}{c}{With $ V^{\urm{rea}} $} & \multicolumn{2}{c}{Without $ V^{\urm{rea}} $} \\
                                  & & & \multicolumn{1}{c}{$ f = -2.032 $} & \multicolumn{1}{c}{$ f = -2.032 $} & \multicolumn{1}{c}{$ f = -2.172 $} \\
      \hline 
      Neutron & 
                $ 1f_{7/2} $               & QP energy         &    2.1610 &     1.5579 &     2.1831 \\
                                  &        & $ v^2 $           &    0.4060 &     0.4423 &     0.4020 \\
                                  &        & QP radius ($ v $) &    4.1214 &     4.0627 &     4.0889 \\
      \hline
      Proton & 
               $ 1s_{1/2} $                & SP energy &  -32.7714 &   -33.5361 &   -33.4290 \\
                                  &        & SP radius &    2.6235 &     2.6335 &     2.6347 \\
      \hline
      Proton & 
               $ 1p_{3/2} $                & SP energy &  -24.7256 &   -25.5808 &   -25.4739 \\
                                  &        & SP radius &    3.2176 &     3.2034 &     3.2099 \\
      \hline
      Proton & 
               $ 1p_{1/2} $                & SP energy &  -21.9096 &   -22.7621 &   -22.6624 \\
                                  &        & SP radius &    3.1808 &     3.1519 &     3.1636 \\
      \hline
      Proton & 
               $ 1d_{5/2} $                & SP energy &  -15.7724 &   -16.5902 &   -16.5052 \\
                                  &        & SP radius &    3.6790 &     3.6417 &     3.6530 \\
      \hline
      Proton & 
               $ 2s_{1/2} $                & SP energy &  -11.7425 &   -12.0260 &   -12.0800  \\
                                  &        & SP radius &    3.6690 &     3.6036 &     3.6202  \\
      \hline
      Proton & 
               $ 1d_{3/2} $                & SP energy &  -10.6618 &   -11.2974 &   -11.2868  \\
                                  &        & SP radius &    3.7202 &     3.6607 &     3.6786  \\
      \hline
      \multicolumn{3}{l}{Proton radius}       &    3.4562 &     3.4200 &     3.4316 \\
      \multicolumn{3}{l}{Neutron radius}      &    3.5525 &     3.5063 &     3.5307 \\
      \multicolumn{3}{l}{Neutron pairing gap} &    1.4978 &     1.0505 &     1.4985 \\
    \end{tabular}
  \end{ruledtabular}
\end{table}
\endgroup
\par  
Interestingly, these changes can also be obtained without introducing $ V^{\urm{rea}} $,
but by enhancing the pairing strength $ f $ from $ f = -2.032 $ to $ f = -2.172 $,
so that the value of the neutron pairing gap of $ \nuc{Ca}{44}{} $,
$ \Delta_n \approx 1.5 \, \mathrm{MeV} $ [see Eq.~\eqref{eq:gap_ref_Ca}] is reproduced
(see the last column of Table~\ref{tab:FaNDF0_020_044_proton}).
The values of $ \delta^2 $ of $ \mathrm{Ca} $ isotopes
(see Fig.~\ref{fig:FaNDF0_refit_Rch_020})
and of the charge density in $ \nuc{Ca}{44}{} $ [see Figs.~\ref{fig:FaNDF0_charge-den_020_044_refit}(a) and \ref{fig:FaNDF0_charge-den_020_044_refit}(b)]
calculated with $ f = -2.172 $ and without $ V^{\urm{rea}} $ lie about halfway between those obtained excluding $ V^{\urm{rea}} $ and those obtained including it,
when $ f = -2.032 $.
An exception is the case of $ \nuc{Ca}{48}{} $, which loses its magic character.
The $ \mathrm{Sn} $ and $ \mathrm{Pb} $ isotopes also show a similar behavior,
while $ \nuc{Sn}{132}{} $ and $ \nuc{Pb}{208}{} $ obtained without the rearrangement potential with the refitted strength
are no longer doubly magic.
\par
One can conclude that the increase of charge radius is indeed a specific effect of the rearrangement potential.
This change in the 
mean-field potential makes the proton single-particle orbitals in $ \nuc{Ca}{44}{} $ less bound,
so that the proton radius calculated with $ f = -2.032 $ increases from 
$ 3.420 \, \mathrm{fm} $ without $ V^{\urm{rea}} $ to $ 3.456 \, \mathrm{fm} $ with it.
Also, the neutron radius increases from $ 3.506 $ to $ 3.552 \, \mathrm{fm} $. 
The radii of the proton single-particle orbitals
are about $ 0.01 $--$ 0.07 \, \mathrm{fm} $ larger
than those without the rearrangement potential,
except for the $ 1s_{1/2} $ orbital.
The radii of the $ sd $-shell orbitals increase more than those of the $ p $-shell orbitals.
\par
If the overall strength of the pairing potential is refitted without the rearrangement potential,
one obtains a much more modest increase of the 
proton radius, from $ 3.420 \, \mathrm{fm} $ to $ 3.432 \, \mathrm{fm} $.
The radii of the proton single-particle orbitals increase at most by $ 0.02 \, \mathrm{fm} $.
The neutron radius also increases from $ 3.506 \, \mathrm{fm} $ to $ 3.531 \, \mathrm{fm} $.
\par
We note that $ V^{\urm{rea}} $ is isoscalar
and therefore both proton and neutron radii increase with finite values of $ \tilde{h}_0 $ and/or $ \tilde{h}_{\urm{D}} $.
The neutron-skin thickness $ \Delta R_{np} = R_n - R_p $ is almost unchanged due to the rearrangement potential (see Table~\ref{tab:FaNDF0_020_044_proton}).
\subsubsection{Brief summary about the rearrangement potential}
\par
Summarizing, the strongest rearrangement potentials are obtained by adopting large values of $ \tilde{h}_0 $ and $ \tilde{h}_{\urm{D}}$.
Such a $ V^{\urm{rea}} $ gives a repulsive contribution to the neutron and proton mean fields in the interior of the nucleus
and an attractive contribution of smaller magnitude at its surface.
This contribution is accompanied by an increase of the neutron pairing field and pairing gap
and by a modification of the occupation probabilities and quasiparticle energies.
The proton single-particle levels become less bound, as compared to a calculation without $ V^{\urm{rea}} $,
and some part of the density is moved from the interior to the exterior of the nucleus, increasing the neutron and proton radii.
\par
To some extent, the action of $ V^{\urm{rea}} $ on the pairing properties can be simulated  by increasing 
the pairing strength $ f $, so as to keep the neutron pairing gap unchanged when $ V^{\urm{rea}} $ is neglected.
By doing this, the charge radii still increase, but significantly less than by including $ V^{\urm{rea}} $.
%
%
\section{Summary}
\label{sec:summary}
\par
We have studied in detail the particle-particle ($ p $-$ p $) channel of {\FaNDF},
the zero-range version of the Fayans EDF.
We have carried out HFB calculations with an energy cutoff $ E_{\urm{cut}} = 60 \, \mathrm{MeV} $,
reducing the pairing strength by $ 15 \, \% $ in the case of protons,
taking into account the Coulomb antipairing effect.
The $ p $-$ p $ channel contains three parameters,
namely, the overall pairing strength $ f $ and the coefficients $ \tilde{h}_0 $ and $ \tilde{h}_{\urm{D}} $
of the density-dependent term and of the density-gradient term.
We have studied $ 25 $ parametrizations, associated with different values of $ \tilde{h}_0 $ and $ \tilde{h}_{\urm{D}} $.
We have found that it is possible to determine a set of values for the three constants,
in such a way as to reproduce the pairing gaps derived from the odd-even
experimental mass differences in $ \nuc{Ca}{44}{} $ accurately,
as well as $ \nuc{Sn}{120}{} $ and $ \nuc{Pb}{204}{} $ within a precision of $ 250 \, \mathrm{keV} $,
ensuring the doubly magic character of $ \nuc{Ca}{40}{} $, $ \nuc{Sn}{132}{} $, and $ \nuc{Pb}{208}{} $ simultaneously.
We have then selected four representative sets
$ \left( \tilde{h}_0, \tilde{h}_{\urm{D}} \right) = \left( 0.75, 0.00 \right) $,
$ \left( 0.75, 0.25 \right) $,
$ \left( 1.00, 0.75 \right) $, and
$ \left( 1.00, 1.00 \right) $,
which yield satisfactory values of the gaps along the three isotopic chains.
Each $ \left( \tilde{h}_0, \tilde{h}_{\urm{D}} \right) $ set is complemented by the pairing strength $ f $ reported in Table~\ref{tab:strength_FaNDF0}.
The value of $ f $ is mostly determined by the parameter $ \tilde{h}_0 $,
which, in turn, dominates the pairing potential [see Eq.~\eqref{eq:pot_pp}].
\par
While these four sets  produce reasonable values of the neutron gaps
for $ \mathrm{Ca} $, $ \mathrm{Sn} $, and $ \mathrm{Pb} $ isotopes in the range of $ 40 < A < 48 $, $ 100 < A < 132 $, and $ 184 < A < 208 $, respectively,
the two sets $ \left( 0.75, 0.00 \right) $ and $ \left( 0.75, 0.25 \right) $ give reasonable overall results compared with experiment also for the charge radii,
with the exception of $ \mathrm{Ca} $ isotopes for $ 40 < A < 48 $.
In this case, the pronounced bell shape experimentally observed is only reproduced by the sets $ \left( 1.00, 0.75 \right) $ and $ \left( 1.00, 1.00 \right) $.
However, these two sets produce neutron gaps that increase too rapidly in nuclei heavier than the magic nuclei $ \nuc{Ca}{48}{} $, $ \nuc{Sn}{132}{} $, and $ \nuc{Pb}{208}{} $.
These parameter sets also produce arch shapes in the charge radii of $ \mathrm{Sn} $ and $ \mathrm{Pb} $ chains which are too pronounced compared to experiment.
Thus, in the present study, we have not been able to find an optimal form of the Fayans pairing
that can reproduce radii equally well in all three isotopic chains.
\par
We have examined in detail the rearrangement potential $ V^{\urm{rea}} $,
confirming that this quantity plays a crucial role in the isotopic dependence of the charge radii.
We found that both $ \tilde{h}_0 $ and $ \tilde{h}_{\urm{D}} $ contribute to the repulsive nature of the rearrangement potential.
On top of it, it was shown that the overall strength of the pairing interaction $ \left| f \right| $ is larger if $ \tilde{h}_0 $ is larger.
Consequently, larger $ \tilde{h}_0 $ and $ \tilde{h}_{\urm{D}} $ give larger charge radii.
Note that even the standard density-dependent pairing interaction ($ \tilde{h}_{\urm{D}} = 0.00 $),
which has been considered for a long time, produces a rearrangement potential.
It is, however, too weak to reproduce the experimental value of 
the charge radius of $ \nuc{Ca}{44}{} $. 
In other words, the role of the rearrangement 
potential in shaping the isotopic trend of the charge radii is, to a large extent, unique:
If we try to mock it up by a refit of the pairing strength, we cannot account for the experimental data quantitatively.
\par
We have also found that
such a repulsive rearrangement potential hardly changes the occupation probabilities but changes the single-particle or quasiparticle energies.
Accordingly, the radii of the single-particle orbitals increase and, as a result, the root-mean-square radii of densities also increase.
\par
Our work has investigated the pairing part of the {\FaNDF} EDF. 
We have not tried to refit the particle-hole ($ p $-$ h $) part of this EDF, and the 
shortcomings we have found must be attributed specifically to {\FaNDF}.
In a step-by-step strategy, one should consider if a better description of all
experimental data (including not only the bell shape of $ \mathrm{Ca} $ isotopes but also the radii in the other isotopes)
can be achieved by refitting the $ p $-$ h $ part of the EDF,
possibly using modern optimization techniques, such as the Bayesian technique~\cite{
  Klausner2025Phys.Rev.C111_014311}.
One could also move to more general forms of the $ p $-$ h $ and of the pairing part.
This endeavor has already been started in the
current literature, e.g.,~by suggesting a more general isovector pairing interaction~\cite{
  Reinhard2024J.Phys.G51_105101}
or a finite-range interaction~\cite{
  Lalit2026Phys.Rev.C113_054310}.
%
%
\begin{acknowledgments}
  The authors thank
  Mario Centelles,
  Nobuo Hinohara,
  Javier Men\'{e}ndez, 
  Takayuki Miyagi, 
  Hitoshi Nakada,
  Alfredo Poves,
  Mengying Qiu,
  and
  Paul-Gerhard Reinhard
  for the fruitful discussion on the Fayans pairing interaction.
  T.N.~acknowledges
  the JSPS Grant-in-Aid for Transformative Research Areas (A) under Grant No.~JP25H01558,
  the JSPS Grant-in-Aid for Scientific Research (S) under Grant No.~JP25K24695,
  the JSPS Grant-in-Aid for Scientific Research (B) under Grant No.~JP25K01003,
  the JSPS Grant-in-Aid for Early-Career Scientists under Grant No.~JP24K17057,
  the JSPS Grant-in-Aid for JSPS Fellows under Grant No.~JP25KJ0405,
  and
  JST COI-NEXT Grant No.~JPMJPF2221.
  X.R.-M. acknowledges support by
  MI-CIU/AEI/10.13039/501100011033 
  and
  by FEDER UE through grants No.~PID2023-147112NB-C22 and No.~CNS2025-165430;
  and through the ``Unit of Excellence Mar\'{\i}a de Maeztu 2025--2028''
  award to the Institute of Cosmos Sciences, grant No.~CEX2024-001451-M.
  Additional support is provided by the Generalitat de Catalunya (AGAUR) through grant No.~2021SGR01095.
  H.S.~acknowledges the JSPS Grant-in-Aid for Scientific Research (C) under Grant No.~JP26K07079. 
  The numerical calculations were performed on cluster computers at the RIKEN iTHEMS Center.
\end{acknowledgments}
\appendix
%
\section{Results with other $ p $-$ h $ EDFs}
\label{sec:Fy-Dr_SkMs}
\par
\begin{figure}[tb]
  \centering
  \includegraphics[width=1.0\linewidth]{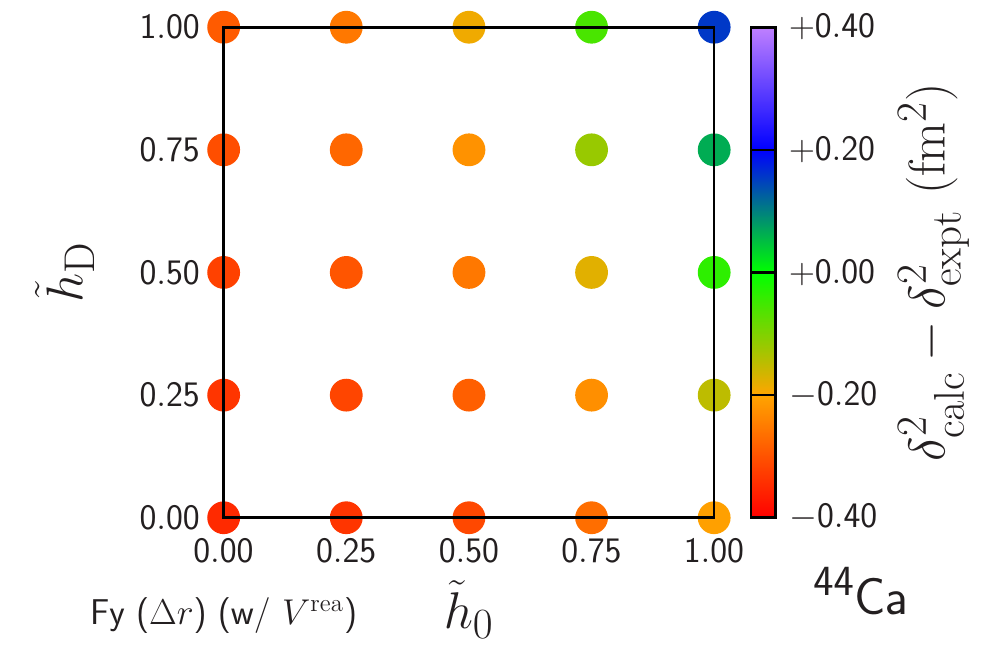}
  \caption{Same as Fig.~\ref{fig:FaNDF0_2D_Rch_020_044} but with Fy ($ \Delta r $) EDF.}
  \label{fig:Fy-Dr_2D_Rch_020_044}
\end{figure}
\begin{figure}[tb]
  \centering
  \includegraphics[width=1.0\linewidth]{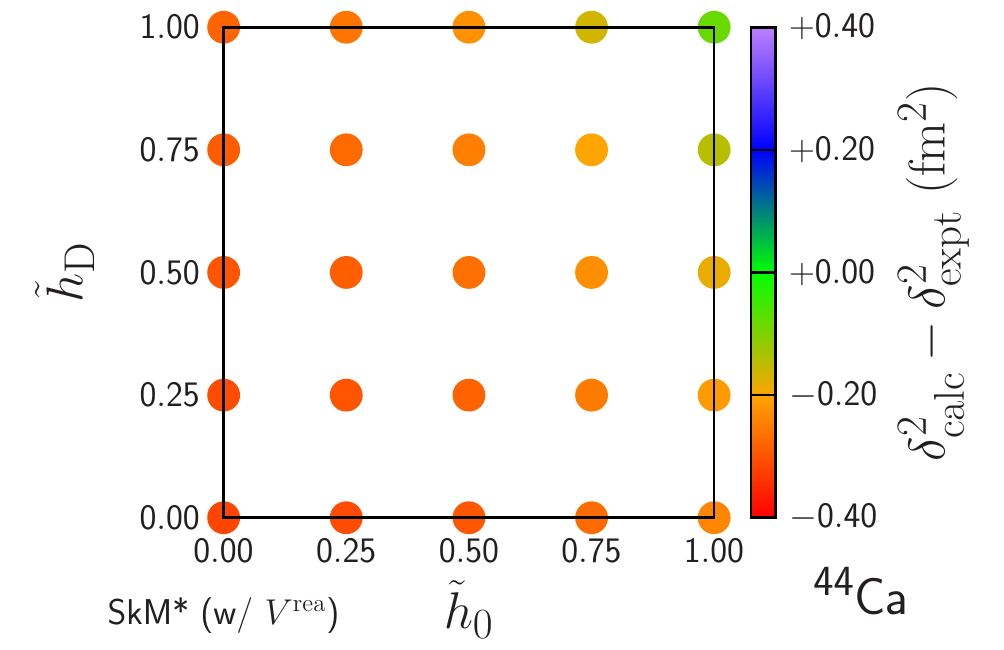}
  \caption{Same as Fig.~\ref{fig:FaNDF0_2D_Rch_020_044} but with SkM* EDF.}
  \label{fig:SkMs_2D_Rch_020_044}
\end{figure}

To demonstrate that the main arguments of the paper hold also in the case of other EDFs,
we show the results with Fy ($ \Delta r $) and SkM* EDFs.
First, the pairing strengths determined as we did for the {\FaNDF} are summarized in Tables~\ref{tab:strength_Fy-Dr} and \ref{tab:strength_SkMs}.
Figures~\ref{fig:Fy-Dr_2D_Rch_020_044} and \ref{fig:SkMs_2D_Rch_020_044} show the $ \delta^2 $ of $ \nuc{Ca}{44}{} $
obtained by the Fy ($ \Delta r $) and SkM* EDFs, respectively.
Different pairs of $ \tilde{h}_0 $ and $ \tilde{h}_{\urm{D}} $ are suitable to reproduce the experimental $ \delta^2 $
for different EDFs;
however, the qualitative discussion remains the same as in the {\FaNDF} case.
\par
\begin{figure}[tb]
  \centering
  \includegraphics[width=1.0\linewidth]{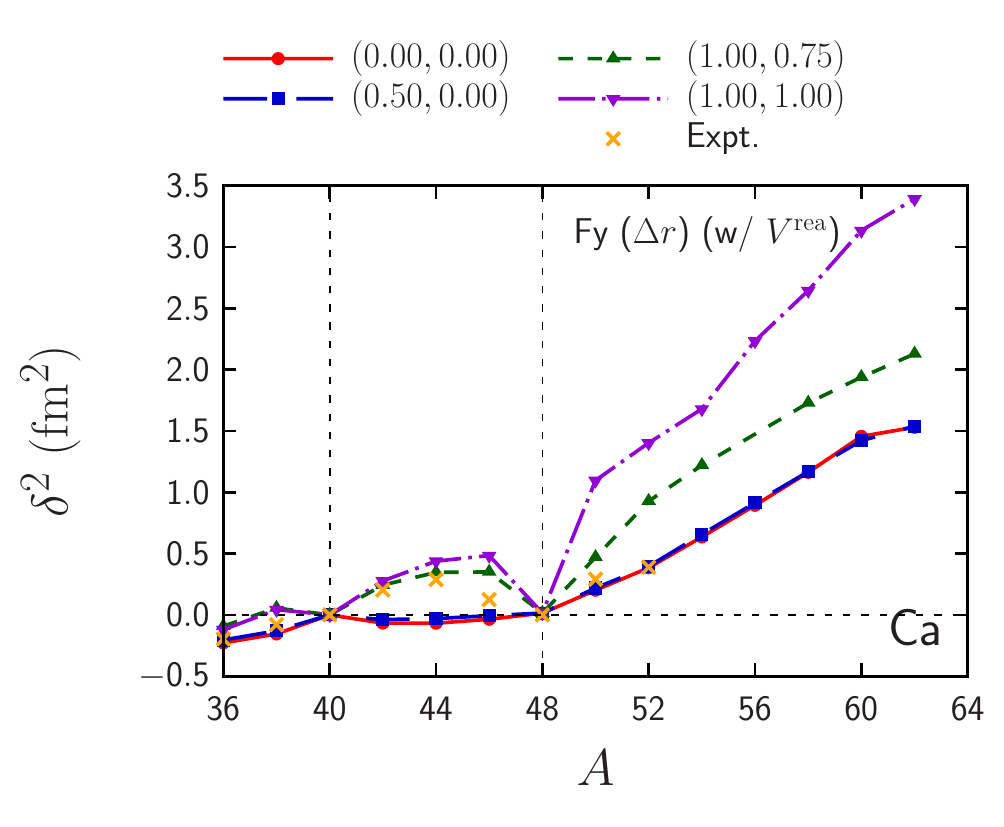}
  \caption{Same as Fig.~\ref{fig:FaNDF0_Rch_020} but with Fy ($ \Delta r $) EDF.}
  \label{fig:Fy-Dr_Rch_020}
\end{figure}
\begin{figure}[tb]
  \centering
  \includegraphics[width=1.0\linewidth]{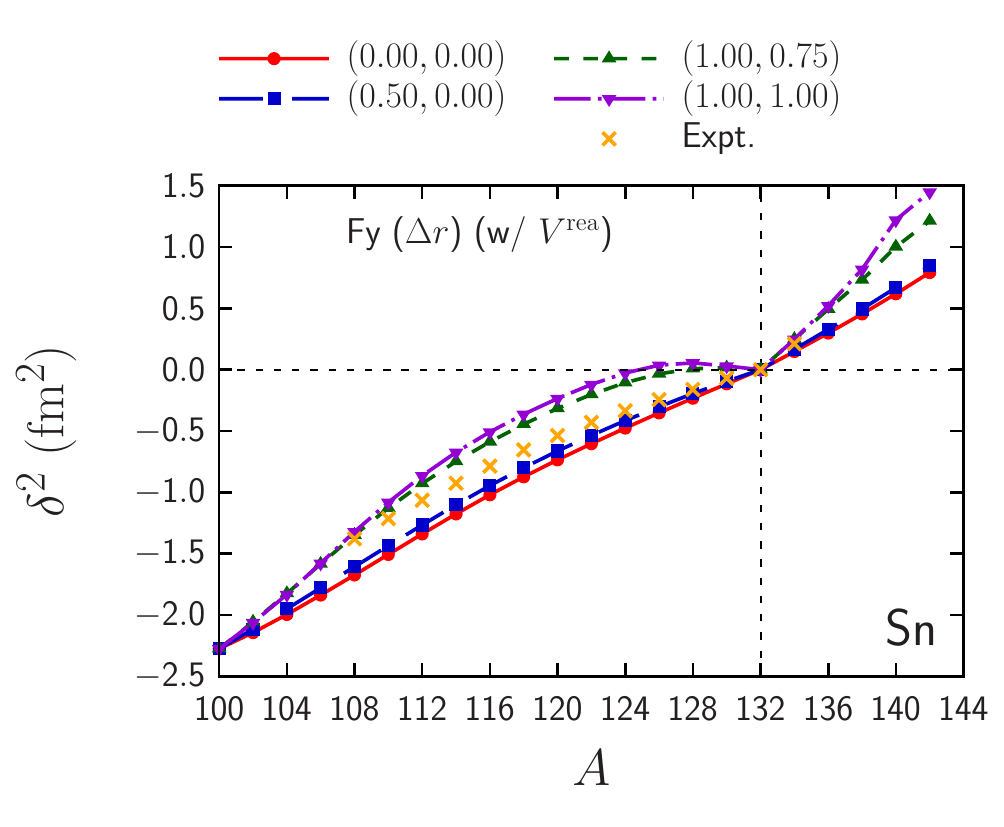}
  \caption{Same as Fig.~\ref{fig:Fy-Dr_Rch_020} but for $ \mathrm{Sn} $ isotopes.}
  \label{fig:Fy-Dr_Rch_050}
\end{figure}
\begin{figure}[tb]
  \centering
  \includegraphics[width=1.0\linewidth]{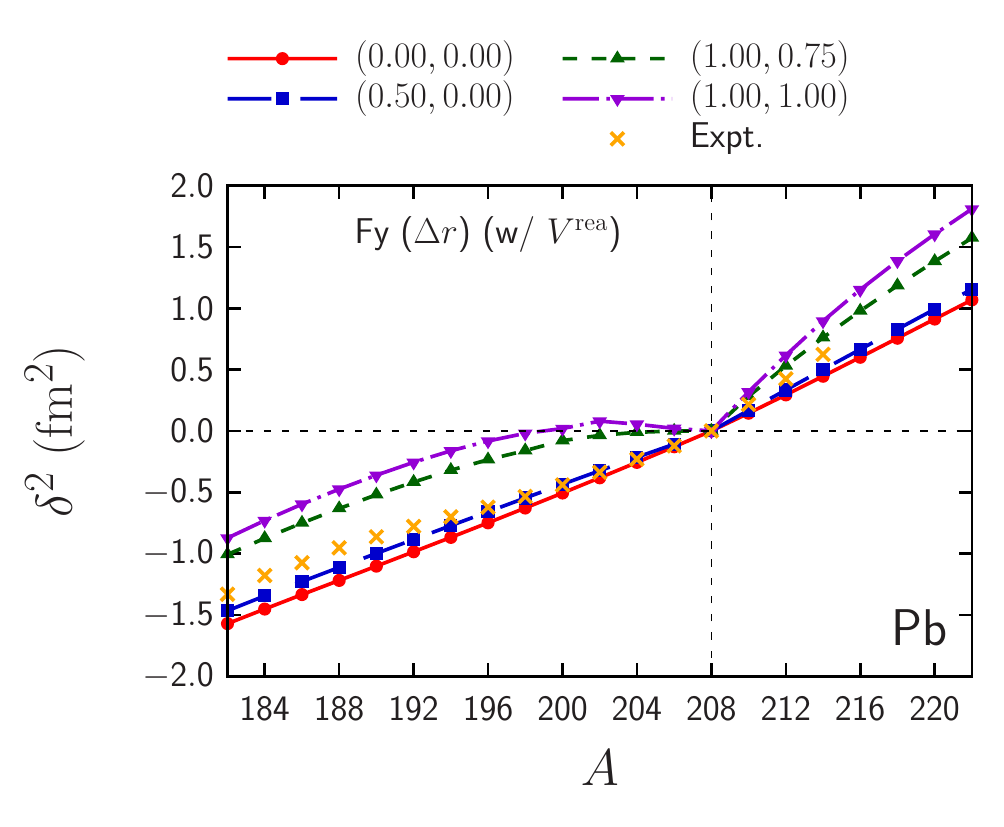}
  \caption{Same as Fig.~\ref{fig:Fy-Dr_Rch_020} but for $ \mathrm{Pb} $ isotopes.}
  \label{fig:Fy-Dr_Rch_082}
\end{figure}

We have tried, as in the case of {\FaNDF}, to single out pairing strengths satisfying all the conditions;
however, no such choice exists for SkM* and, thus, we only consider the Fy ($ \Delta r $) EDF.
Figures~\ref{fig:Fy-Dr_Rch_020}, \ref{fig:Fy-Dr_Rch_050}, and \ref{fig:Fy-Dr_Rch_082} show the mass number $ A $ dependence of $ \delta^2 $
for $ \mathrm{Ca} $, $ \mathrm{Sn} $, and $ \mathrm{Pb} $ isotopes, respectively.
As in the case of {\FaNDF}, the pairing interactions with larger $ \tilde{h}_0 $ and $ \tilde{h}_{\urm{D}} $ reproduce
the bell-shaped behavior of $ \mathrm{Ca} $ isotopes,
while they do not work well in reproducing the data for the $ \mathrm{Sn} $ and $ \mathrm{Pb} $ isotopes.
\begin{table}[tb]
  \centering
  \caption{Same as Table~\ref{tab:strength_FaNDF0} but with the Fy ($ \Delta r $) EDF.}
  \label{tab:strength_Fy-Dr}
  \begin{ruledtabular}
    \begin{tabular}{dddcccccl}
      \multicolumn{1}{c}{$ \tilde{h}_0 $} & \multicolumn{1}{c}{$ \tilde{h}_{\urm{D}} $} & \multicolumn{1}{c}{$ -f $} & \multicolumn{1}{c}{$ \nuc{Sn}{120}{} $} & \multicolumn{1}{c}{$ \nuc{Pb}{204}{} $} & \multicolumn{1}{c}{$ \nuc{Ca}{40}{} $} & \multicolumn{1}{c}{$ \nuc{Sn}{132}{} $} & \multicolumn{1}{c}{$ \nuc{Pb}{208}{} $} & \\
      \hline
      0.00 & 0.00 & 0.493 &            &            & \checkmark & \checkmark & \checkmark &  \\
      0.00 & 0.25 & 0.516 &            &            & \checkmark & \checkmark & \checkmark &  \\
      0.00 & 0.50 & 0.540 &            &            & \checkmark & \checkmark & \checkmark &  \\
      0.00 & 0.75 & 0.566 &            &            & \checkmark & \checkmark & \checkmark &  \\
      0.00 & 1.00 & 0.592 &            &            & \checkmark & \checkmark &            &  \\
      \hline
      0.25 & 0.00 & 0.606 &            &            & \checkmark & \checkmark & \checkmark &  \\
      0.25 & 0.25 & 0.642 &            &            & \checkmark & \checkmark & \checkmark &  \\
      0.25 & 0.50 & 0.680 &            &            & \checkmark & \checkmark & \checkmark &  \\
      0.25 & 0.75 & 0.721 &            &            & \checkmark & \checkmark & \checkmark &  \\
      0.25 & 1.00 & 0.764 &            &            & \checkmark & \checkmark &            &  \\
      \hline
      0.50 & 0.00 & 0.779 & \checkmark & \checkmark & \checkmark & \checkmark & \checkmark & $ \star $ \\
      0.50 & 0.25 & 0.839 &            &            & \checkmark & \checkmark & \checkmark &  \\
      0.50 & 0.50 & 0.905 &            &            & \checkmark & \checkmark & \checkmark &  \\
      0.50 & 0.75 & 0.975 &            &            & \checkmark & \checkmark & \checkmark &  \\
      0.50 & 1.00 & 1.052 &            &            & \checkmark & \checkmark &            &  \\
      \hline
      0.75 & 0.00 & 1.057 & \checkmark & \checkmark &            & \checkmark & \checkmark &  \\
      0.75 & 0.25 & 1.163 & \checkmark & \checkmark &            & \checkmark & \checkmark &  \\
      0.75 & 0.50 & 1.278 &            &            &            & \checkmark & \checkmark &  \\
      0.75 & 0.75 & 1.403 &            &            & \checkmark & \checkmark & \checkmark &  \\
      0.75 & 1.00 & 1.535 &            &            & \checkmark & \checkmark &            &  \\
      \hline
      1.00 & 0.00 & 1.487 & \checkmark & \checkmark &            & \checkmark & \checkmark &  \\
      1.00 & 0.25 & 1.647 & \checkmark & \checkmark &            & \checkmark & \checkmark &  \\
      1.00 & 0.50 & 1.805 & \checkmark & \checkmark &            & \checkmark & \checkmark &  \\
      1.00 & 0.75 & 1.955 & \checkmark & \checkmark & \checkmark & \checkmark & \checkmark & $ \star $ \\
      1.00 & 1.00 & 2.092 & \checkmark & \checkmark & \checkmark & \checkmark & \checkmark & $ \star $ \\
    \end{tabular}
  \end{ruledtabular}
\end{table}
\begin{table}[tb]
  \centering
  \caption{Same as Table~\ref{tab:strength_FaNDF0} but with the SkM* EDF.}
  \label{tab:strength_SkMs}
  \begin{ruledtabular}
    \begin{tabular}{dddcccccl}
      \multicolumn{1}{c}{$ \tilde{h}_0 $} & \multicolumn{1}{c}{$ \tilde{h}_{\urm{D}} $} & \multicolumn{1}{c}{$ -f $} & \multicolumn{1}{c}{$ \nuc{Sn}{120}{} $} & \multicolumn{1}{c}{$ \nuc{Pb}{204}{} $} & \multicolumn{1}{c}{$ \nuc{Ca}{40}{} $} & \multicolumn{1}{c}{$ \nuc{Sn}{132}{} $} & \multicolumn{1}{c}{$ \nuc{Pb}{208}{} $} & \\
      \hline
      0.00 & 0.00 & 0.579 &            &            & \checkmark & \checkmark & \checkmark &  \\
      0.00 & 0.25 & 0.602 &            &            & \checkmark & \checkmark & \checkmark &  \\
      0.00 & 0.50 & 0.626 &            &            & \checkmark & \checkmark & \checkmark &  \\
      0.00 & 0.75 & 0.651 &            &            & \checkmark & \checkmark & \checkmark &  \\
      0.00 & 1.00 & 0.678 &            &            & \checkmark & \checkmark & \checkmark &  \\
      \hline
      0.25 & 0.00 & 0.696 &            & \checkmark & \checkmark & \checkmark & \checkmark &  \\
      0.25 & 0.25 & 0.729 &            &            & \checkmark & \checkmark & \checkmark &  \\
      0.25 & 0.50 & 0.764 &            &            & \checkmark & \checkmark & \checkmark &  \\
      0.25 & 0.75 & 0.802 &            &            & \checkmark & \checkmark & \checkmark &  \\
      0.25 & 1.00 & 0.843 &            &            & \checkmark & \checkmark & \checkmark &  \\
      \hline
      0.50 & 0.00 & 0.864 &            & \checkmark & \checkmark & \checkmark & \checkmark &  \\
      0.50 & 0.25 & 0.915 &            & \checkmark & \checkmark & \checkmark & \checkmark &  \\
      0.50 & 0.50 & 0.970 &            &            & \checkmark & \checkmark & \checkmark &  \\
      0.50 & 0.75 & 1.031 &            &            & \checkmark & \checkmark & \checkmark &  \\
      0.50 & 1.00 & 1.097 &            &            & \checkmark & \checkmark & \checkmark &  \\
      \hline
      0.75 & 0.00 & 1.115 & \checkmark & \checkmark &            & \checkmark & \checkmark &  \\
      0.75 & 0.25 & 1.197 & \checkmark & \checkmark &            & \checkmark & \checkmark &  \\
      0.75 & 0.50 & 1.287 &            & \checkmark &            & \checkmark & \checkmark &  \\
      0.75 & 0.75 & 1.385 &            & \checkmark &            & \checkmark & \checkmark &  \\
      0.75 & 1.00 & 1.490 &            &            &            & \checkmark & \checkmark &  \\
      \hline
      1.00 & 0.00 & 1.483 & \checkmark & \checkmark &            & \checkmark & \checkmark &  \\
      1.00 & 0.25 & 1.605 & \checkmark & \checkmark &            & \checkmark & \checkmark &  \\
      1.00 & 0.50 & 1.733 & \checkmark & \checkmark &            & \checkmark & \checkmark &  \\
      1.00 & 0.75 & 1.863 & \checkmark & \checkmark &            & \checkmark & \checkmark &  \\
      1.00 & 1.00 & 1.990 & \checkmark & \checkmark &            & \checkmark & \checkmark &  \\
    \end{tabular}
  \end{ruledtabular}
\end{table}
%
%
%
%
%
\end{document}